\documentclass[11pt,a4paper]{article}
\pdfoutput=1
\usepackage{jcappub}
\usepackage{amsmath}
\usepackage{booktabs}
\usepackage{array}
\usepackage{xcolor}
\definecolor{stepN}{RGB}{0,159,223}
\definecolor{stepP}{RGB}{241,143,31}
\definecolor{stepG}{RGB}{208,0,111}
\definecolor{stepB}{RGB}{0,168,62}
\definecolor{UUred}{HTML}{990000}
\hypersetup{linkcolor=UUred, citecolor=UUred, urlcolor=UUred}
\DeclareRobustCommand{\alt}[2]{\hyperlink{alt:#1#2}{\textcolor{step#1}{#1#2}}}
\DeclareRobustCommand{\lk}[1]{\hyperlink{alt:#1}{#1}}
\DeclareRobustCommand{\altanchor}[1]{\hypertarget{alt:#1}{}}
\usepackage{xspace}
\usepackage{placeins}
\usepackage{enumitem}
\usepackage{tikz}
\newcommand{\bubblebullet}{\tikz[baseline=-0.55ex, x=1.15em, y=1.15em, line width=0.6pt]{%
  \draw[UUred] (-0.23,-0.02) circle (0.17); \draw[UUred] (0.13,0.08) circle (0.23);
  \draw[UUred] (0.30,-0.18) circle (0.11);}}

\newcommand{\TL}{\texttt{Transition\-Listener}\xspace}
\newcommand{\DRalgo}{\texttt{DRalgo}\xspace}
\newcommand{\Tp}{T_{\rm p}}
\newcommand{\Treh}{T_{\rm reh}}
\newcommand{\Adyn}{A_{\rm dyn}}
\newcommand{\Astat}{A_{\rm stat}}
\newcommand{\muz}{\mu_0}
\newcommand{\OGW}{\Omega_{\rm GW}}
\newcommand{\Neff}{N_{\rm eff}}
\newcommand{\Sthree}{S_3}
\newcommand{\bH}{\beta/H}
\newcommand{\bHR}{(\beta/H)_R}
\newcommand{\bHS}{(\beta/H)_S}
\newcommand{\bHG}{(\beta/H)_\Gamma}
\newcommand{\eu}{\mathrm{e}}
\newcommand{\dd}{\mathrm{d}}

\newcommand{\nb}{n_{\rm b}}
\newcommand{\cs}{c_{\rm s}}
\newcommand{\HV}{H_{\rm V}}
\newcommand{\fperc}{f_{\rm perc}}
\newcommand{\DM}{\mathcal{D}}

\title{Hubble, bubble, toil and trouble: Uncertainties in the nanohertz signal of a classically conformal
dark \texorpdfstring{$U(1)$}{U(1)}}

\author{Carlo Tasillo}
\affiliation{Instituto de F\'isica Corpuscular (IFIC), Universitat de
Val\`encia--CSIC,\\ C/ Catedr\'atico Jos\'e Beltr\'an 2, E-46980 Paterna, Spain}
\emailAdd{carlo.tasillo@ific.uv.es}

\abstract{Pulsar timing arrays have found evidence for a nanohertz gravitational wave background which may
partly originate from a strongly supercooled phase transition in a classically conformal dark $U(1)$ sector.
Combining the next-to-leading-order nucleation rate from the three-dimensional effective field theory with the most
accurate treatment available for each subsequent step of the signal prediction, we infer the properties of
such a sector from the data.
For the NANOGrav 15\,yr data, we find a gauge coupling $g = 0.43$ to $0.50$, dark photon masses
around a GeV, and reheating temperatures around the $100$\,MeV scale.
Repeating the fit for $20$ alternative modelling and data choices, including IPTA DR2, shows that the inferred energy
scale is most sensitive to the choice of the gravitational wave spectral templates.
Across templates for highly boosted fluid shells, the $1\sigma$ ranges of the dark Higgs vacuum expectation value $v$
span more than an order of magnitude, and more than two with sound-wave sources.
The preferred region moves by two to three standard deviations when the
modelling of dispersion measure variations in pulsar timing is varied, when 
the instantaneous reheating approximation is used,
or when the expected astrophysical binary background is included.
If these binaries need more than about $5$\,Gyr to harden, the phase-transition contribution is preferred at
more than $3\sigma$ under astrophysical priors from recent galaxy surveys.
The entire analysis can be reproduced with the public codes \texttt{TransitionListener} and \texttt{binaryfog}.}


\begin{document}
\maketitle
\flushbottom

\section{Introduction}
\label{sec:intro}

Several pulsar timing array (PTA) collaborations have reported evidence for a stochastic gravitational
wave (GW) background at nanohertz frequencies. After a common red noise spectrum was identified in earlier
data~\cite{NANOGrav:2020bcs, Goncharov:2021oub, EPTA:2021crs, Perera:2019sca, Antoniadis:2022pcn},
the analyses of NANOGrav, EPTA, PPTA, and CPTA
found the Hellings--Downs correlation~\cite{Hellings:1983fr} at about $2\sigma$ to
$4\sigma$~\cite{NANOGrav:2023gor, EPTA:2023fyk, Reardon:2023gzh, Xu:2023wog}, with
spectra that a comparison by the IPTA found
consistent~\cite{NANOGrav:2023hde, EPTA:2023sfo, Tarafdar:2022toa, Zic:2023gta,InternationalPulsarTimingArray:2023mzf}.
The signal has since been reported independently by MeerKAT~\cite{Miles:2024rjc,
Miles:2024seg}, InPTA~\cite{Rana:2025ano, Tahbildar:2026sqj}, and in a combined
search~\cite{Yu:2025dor}. Inspiralling supermassive black hole binaries are the expected
source~\cite{Kelley:2016gse, Kelley:2017lek, Casey-Clyde:2021xro, Ellis:2023dgf, Depta:2023uhy}, but the measured
slope and amplitude of their contribution lie at the upper end of the range predicted by population
models~\cite{NANOGrav:2023hfp,
Sato-Polito:2023gym, Sato-Polito:2025ivz, Chen:2025wel}. The signal may therefore contain an additional
cosmological contribution, from cosmic strings, domain walls, inflation, or scalar-induced
GWs~\cite{NANOGrav:2023hvm, EPTA:2023xxk, Figueroa:2023zhu, Ellis:2023oxs, Madge:2023dxc}, or from a
first-order phase transition in a dark sector around the $100$\,MeV scale~\cite{Nakai:2020oit,
NANOGrav:2021flc, Freese:2022qrl, Bringmann:2023opz, Fujikura:2023lkn, Athron:2023mer,
Gouttenoire:2023bqy, Salvio:2023blb, Winkler:2024olr, Goncalves:2025uwh, Li:2025nja, Costa:2025csj,
Balan:2025uke, Bringmann:2026xcx, Biondini:2026uds, Puchades-Ibanez:2026ksh}.

Phase transitions in classically conformal dark sectors are known to supercool strongly and produce loud GW
signals~\cite{Jinno:2016knw, Iso:2017uuu,
Marzola:2017jzl, Prokopec:2018tnq, Baldes:2018emh, Kierkla:2022odc, Levi:2022bzt, Sagunski:2023ynd,
Gouttenoire:2023pxh, Salvio:2026bco}, making them particularly attractive as they
require the least parameter tuning among other phase-transition
explanations of the PTA signal~\cite{Bringmann:2026xcx}. In the simplest of them, a dark $U(1)$ gauge theory with a dark
Higgs, the signal is parametrised by only two parameters, the gauge coupling $g$ and the vacuum
expectation value $v$ of the dark Higgs~\cite{Balan:2025uke, Puchades-Ibanez:2026ksh}. With the PTA data,
the extraction of model parameters from a measured GW spectrum becomes an inference problem which is
limited by the uncertainties of each step of the prediction
of the GW background. The question of this work is how theoretical, methodological, and
measurement uncertainties impact the model parameter inference and the associated significance
of a phase-transition contribution to the PTA signal.

The theoretical uncertainty of the bubble nucleation rate is the one most widely studied in the
literature~\cite{Croon:2020cgk, Gould:2021oba, Ekstedt:2021kyx, Gould:2021ccf, Athron:2022jyi, Athron:2023xlk,
Kierkla:2023von, Lewicki:2024xan, Gould:2024chm, Kierkla:2025qyz, Christiansen:2025xhv, Puchades-Ibanez:2026ksh,
Kierkla:2026bnm}. In the conventional four-dimensional treatment with daisy resummation the rate depends strongly
on the renormalisation scale, which changes the predicted GW amplitude by orders of magnitude~\cite{Croon:2020cgk,
Gould:2021oba, Athron:2022jyi, Athron:2023xlk, Lewicki:2024xan}. Dimensional reduction instead integrates out the
hard thermal modes and presents a controlled perturbative framework, circumventing this and other issues of
the four-dimensional approach~\cite{Farakos:1994kx, Kajantie:1995dw, Braaten:1995cm, Gould:2021oba,
Ekstedt:2022bff, Ekstedt:2022zro, Gould:2023ovu, Ekstedt:2024etx, Kierkla:2023von}.
The subsequent steps of the signal prediction add further sources of uncertainty, which have so far received little
attention:
\begin{itemize}[label=\bubblebullet, leftmargin=1.6em, itemsep=1pt, parsep=0pt, topsep=3pt]
\item Whether a slow phase transition completes, and when it percolates, depends on how the expansion of the universe
during the phase transition is modelled~\cite{Ellis:2018mja, Athron:2022mmm, Athron:2023mer}.
\item The relation between the bubble separation and the transition speed, setting the peak amplitude and
frequency of the spectrum,
breaks down for slow phase transitions~\cite{Megevand:2016lpr, Matuszak:2026xsz}.
\item The reheating temperature, setting the redshift of the GW signal, differs between
instantaneous and gradual conversion
prescriptions of the liberated vacuum energy~\cite{Matuszak:2026xsz}.
\item None of the GW spectral templates recommended by the LISA cosmology working group has been validated for a phase
transition that is both as slow and as strong as the one considered here~\cite{Caprini:2019egz,
Caprini:2024hue, Caprini:2024gyk, Lewicki:2022pdb}.
\item The free spectra inferred from the PTA data depend on the noise
models of the pulsars and the interstellar medium, as well as the signal's overlap reduction
function~\cite{NANOGrav:2023ctt, Agarwal:2026dlv, Larsen:2026pru}.
\item Supermassive black hole binaries produce a
background of comparable amplitude which depends on both the binary population and the details of
binary hardening~\cite{NANOGrav:2023hfp, holodeck, Bringmann:2026uxo}.
\end{itemize}

\begin{figure}[t]
\centering
\includegraphics[width=\textwidth]{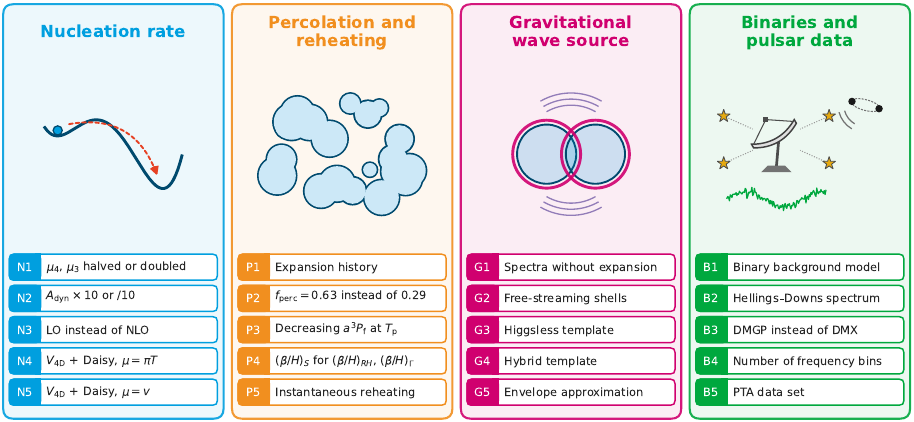}
\caption{The steps of the GW spectral prediction from the nucleation rate to
the pulsar timing likelihood. The smaller boxes indicate the alternative modelling choices discussed
in section~\ref{sec:alternatives}. 
\label{fig:pipeline}}
\end{figure}

Previous studies already compared some of these choices for this model~\cite{Puchades-Ibanez:2026ksh},
for a dark Abelian Higgs model
with a tree-level mass~\cite{Biondini:2026uds}, and for phase transitions at the electroweak
scale~\cite{Guo:2021qcq, Athron:2023rfq}.
In this work we perform the whole computation from the nucleation rate to the PTA likelihood and repeat the
fit for each of the $20$ alternative choices illustrated in
figure~\ref{fig:pipeline}. To isolate the effect of each choice, the reference fit
attributes the whole signal to the phase transition, and the binaries enter as one of the alternatives.

For the NANOGrav data we find that the inference is limited most by the mapping of the thermodynamic
parameters to spectral
templates: the templates for highly boosted fluid shells leave the vacuum expectation value $v$ uncertain at
$1\sigma$ by more
than an order of magnitude, and by more than two orders of magnitude once spectra sourced by sound waves are
included. The noise model of the pulsars, the
share of the binaries, and the reheating prescription each move the preferred region by two to three posterior standard
deviations. The perturbative uncertainties of the next-to-leading-order (NLO)
nucleation rate move the region by less than half a standard deviation, and the leading-order (LO) rate and
the nucleation rate of
the four-dimensional daisy-resummed potential with the couplings run to the thermal scale $\pi T$ by less than one. With
the couplings fixed at the vacuum scale, as in earlier fits of this
model~\cite{Balan:2025uke,Bringmann:2026xcx}, the same rate moves the preferred gauge
coupling $g$ up by about $10\%$.

With the binaries included, the data show
a hint ($1.9\sigma$, wide astrophysical priors) to a preference ($3.6\sigma$, priors at the
widths of single galaxy-survey fits) for a contribution of the phase transition to the observed signal.
If the binaries need more than about $5$\,Gyr to harden, as indicated by the final parsec problem~\cite{Begelman:1980vb,
Milosavljevic:2002bn, Merritt:2004gc}, the significance of a phase-transition
contribution rises up to above $3\sigma$ under priors based on recent galaxy surveys~\cite{Bringmann:2026uxo},
and beyond $4\sigma$
under priors at the width of a single galaxy survey. For an even more inefficient hardening of at least $7$\,Gyr,
the significance reaches between about $3\sigma$ and $5\sigma$, depending on the galaxy survey data used to
infer the binary population.

The whole analysis can be reproduced with two public codes.
\TL~\cite{Matuszak:2026xsz} is extended to allow choosing between the
alternative prescriptions for the nucleation rate, percolation, reheating,
and the GW source through a single setting in its configuration file.
\texttt{binaryfog}~\cite{Bringmann:2026uxo} provides predictions for the astrophysical
contribution to the background based on populations of
supermassive black hole binaries. The scripts that produce every number, table, and
figure of this work are publicly available.
They are archived on Zenodo, 
\begin{center}
\href{https://doi.org/10.5281/zenodo.24516038}{\texttt{doi.org/10.5281/zenodo.24516038}}\\
\end{center}
and will be developed further, including later bug fixes, at
\begin{center}
\href{https://github.com/tasicarl/TransitionListener/tree/uncertainties}{\texttt{github.com/tasicarl/TransitionListener/tree/uncertainties}}\\
\end{center}
so that the analysis can be repeated with other modelling choices and for other models.

This work is organised as follows.
We first predict the signal with the most up-to-date modelling choice at each step of figure~\ref{fig:pipeline}
(section~\ref{sec:signal}), then replace these choices one at a time by the alternatives in use in the literature
and measure how far each moves the inferred parameters, before we discuss the statistical preference for a
phase transition explanation of the PTA data and summarise the results (section~\ref{sec:alternatives}).
Section~\ref{sec:conclusions} concludes this work and draws lessons for future analyses.
The appendices contain technical notes on the computation of the nucleation rate in the EFT, how
the $\mathcal{O}(10^3)$-times faster phase transition predictions in \TL required for the presented analysis
are achieved,
details on the statistical
analysis, the modelling choices with the IPTA DR2 data, and tables for reference.

\section{State-of-the-art prediction of the phase transition signal}
\label{sec:signal}

This section predicts the PTA signal of the dark $U(1)$ and of the supermassive black hole binaries with the modelling
choice at each step of figure~\ref{fig:pipeline} that we consider the state of the art.
After defining each step of the signal prediction in subsections~\ref{sec:model} to~\ref{sec:source}, we
discuss the PTA likelihood and binary backgrounds in subsection~\ref{sec:data} and end this section with the
parameter regions that the data prefer.
The
modelling choices presented in this section are the defaults against which section~\ref{sec:alternatives}
measures the alternatives.

\subsection{The classically conformal dark \texorpdfstring{$U(1)$}{U(1)} model}
\label{sec:model}

The dark sector under consideration is a $U(1)$ gauge theory with gauge coupling $g$ and a charge-one complex scalar
$\Phi = (\phi + \mathrm{i}\chi)/\sqrt2$ with the potential $V = \lambda|\Phi|^4$ and no tree-level mass term, where
$\phi$ is the dark Higgs and $\chi$ the Goldstone mode.
The massive gauge boson of the broken phase is the dark photon $A'$.
We parametrise the model
through two parameters, the renormalisation scale $\muz$ at which the quartic
coupling vanishes in the $\overline{\rm MS}$ scheme, $\lambda(\muz) = 0$,
and the gauge coupling $g \equiv g(\muz)$ at the same scale.
These conditions fix the vacuum expectation value of the dark
Higgs to $v = \eu^{1/6}\muz/g$~\cite{Christiansen:2025xhv}. For an easier interpretation of the parameter space,
we take $g$ and $v$ as the primary parameters of the analysis.
As in earlier analyses of this model~\cite{Balan:2025uke, Goncalves:2025uwh, Costa:2025csj,
Puchades-Ibanez:2026ksh}, we assume that the dark sector is in thermal equilibrium with the Standard Model
(SM) photon bath and decays promptly after the
phase transition~\cite{Bringmann:2023opz, Ellis:2020nnr}, for instance through kinetic mixing of the dark
photon with the SM photon,
couplings to light fermions, a second gauge group, or enhanced couplings between the scalar and electrons. None of these
couplings necessarily changes the emitted GW signal or the cosmological expansion history~\cite{Balan:2025uke,
Bringmann:2026xcx}.

In a strongly supercooled phase transition of a classically conformal model, the critical bubble, the field
configuration that determines the nucleation rate, reaches only field values $\phi$ of the order of $T$, far below
$v$.
The masses of the gauge fields inside the critical bubble are then of order $gT$, and the three-dimensional
EFT of the soft modes applies.
The
vacuum energy and the reheated plasma involve the broken phase at $\phi = v \gg T$, where the dark photon mass
$gv$ exceeds
the temperature and the EFT does \textit{not} apply. We therefore use the hybrid
approach recommended in ref.~\cite{Kierkla:2023von}: we compute the bubble nucleation rate in the EFT of
ref.~\cite{Puchades-Ibanez:2026ksh},
and perform the phase tracing as well as the computation of thermodynamic quantities within each phase using the
four-dimensional one-loop potential with daisy resummation~\cite{Balan:2025uke, Bringmann:2026xcx}.

\subsection{Phases and thermodynamics}
\label{sec:thermo}

\TL\ traces the phases, the minima of the potential, as functions of temperature, and obtains from them
the released vacuum energy and other thermodynamic quantities of the plasma like sound speeds and entropy densities.
All of these follow from the four-dimensional one-loop
potential with daisy resummation of the dark Higgs field $\phi$ at the temperature $T$ in the Arnold--Espinosa
scheme~\cite{Arnold:1992rz, Balan:2025uke, Bringmann:2026xcx},
\begin{multline}
  V(\phi, T) = \frac{\lambda}{4}\phi^4 + \sum_i \frac{n_i\,m_i^4(\phi)}{64\pi^2}
  \left[\ln\frac{m_i^2(\phi)}{\mu^2} - c_i\right]
  + \frac{T^4}{2\pi^2}\sum_i n_i\,J_{\rm B}\!\left(\frac{m_i^2(\phi)}{T^2}\right) \\
  - \frac{T}{12\pi}\sum_i n_i\left[\left(m_i^2(\phi) + \Pi_i\right)^{3/2} - m_i^3(\phi)\right],
  \label{eq:V4D}
\end{multline}
where $J_{\rm B}(x) = \int_0^\infty \dd k\,k^2\ln\big(1 - \eu^{-\sqrt{k^2 + x}}\big)$ is the bosonic thermal function. The
sums run over the dark Higgs and the Goldstone mode ($n_i = 1$ each) and the two transverse ($n_i = 2$) and the
longitudinal ($n_i = 1$) polarisations of the dark photon, with the
field-dependent and thermal masses
\begin{align}
  m_\phi^2 &= 3\lambda\phi^2\,, & m_\chi^2 &= \lambda\phi^2\,, & m_{A'}^2 &= g^2\phi^2\,, \nonumber\\
  \Pi_\phi = \Pi_\chi &= \left(\frac{\lambda}{3} + \frac{g^2}{4}\right)T^2\,, & \Pi_{A',{\rm L}} &= \frac{g^2T^2}{3}\,, &
  \Pi_{A',{\rm T}} &= 0\,.
  \label{eq:masses4D}
\end{align}
In the Landau gauge used here, ghost fields are
massless and do not depend on $\phi$ and therefore do not contribute to the effective potential. In the
thermodynamics, however, they cancel one massless bosonic degree of freedom, so that the symmetric phase holds
the four physical degrees of freedom of the dark sector~\cite{Quiros:1999jp, Laine:2016hma}.
The zero-temperature part in eq.~\eqref{eq:V4D} is the one-loop Coleman--Weinberg potential~\cite{Coleman:1973jx} in the
$\overline{\rm MS}$ scheme at the renormalisation scale $\mu$, with $c_i = 3/2$ for the scalars and $5/6$ for the
dark photon, and the last term is the daisy resummation.

Since every
$m_i^2(\phi)$ is proportional to $\phi^2$, changing $\mu$ or the constants $c_i$ adds only terms proportional to
$\phi^4$, which a shift of $\lambda$ absorbs. Throughout this work we use the renormalisation
conditions of ref.~\cite{Coleman:1973jx} for the potential used within phase tracing.
This choice fixes the derivatives of the potential
at the renormalisation scale $\phi = \mu$.
Moreover, we impose that the minimum at zero temperature lies at $\phi = v$, which fixes the quartic coupling
for a given choice of $\mu$, $v$ and $g$.
We take $\mu = v$, the
scale of the broken minimum, as in refs.~\cite{Balan:2025uke, Goncalves:2025uwh, Bringmann:2026xcx} to keep
the size of the logarithms of the Coleman--Weinberg potential small in phase tracing.
Then
\begin{equation}
  110\,\lambda^2 - 48\pi^2\lambda + 33\,g^4 = 0\qquad \text{and} \qquad
  \Delta V = \frac{3\lambda v^4}{88} \simeq \frac{3g^4v^4}{128\pi^2}\,,
  \label{eq:dV}
\end{equation}
where in the last step we took the smaller root $\lambda \simeq 11g^4/16\pi^2$ of the quadratic equation in order
to express the potential difference $\Delta V$ between both phases, see ref.~\cite{Balan:2025uke} for more details.
In the broken phase the dark photon and the dark Higgs
have the masses
\begin{align}
  m_{A'} &= g\,v\,, &
  m_\phi &= \left(\frac{3}{8\pi^2}\right)^{1/2} g^2 v \simeq 0.2\,g^2 v\,,
  \label{eq:masses}
\end{align}
the latter taken from the curvature of the full one-loop potential at $\phi = v$.

There is an alternative choice of renormalisation conditions used in variants \lk{N4} and \lk{N5} which we
already want to introduce here.
In these variants, the renormalisation conditions described in the previous paragraph are not imposed.
Instead, the couplings are run from $\muz$ to
$\mu$ using the one-loop $\beta$-functions of the model, with $\mu = \pi T$ in the one and $\mu = v$ in the
other case, but
\emph{only} for the computation of the nucleation rate.
Also in these variants, the phase tracing and the thermodynamics are
still computed with the couplings fixed at $\mu = v$ as discussed above.\footnote{
In refs.~\cite{Kierkla:2022odc, Kierkla:2023von, Kierkla:2026bnm} yet another renormalisation
  scale, $\mu = \max(g\phi, \pi T)$, is used to interpolate between these two regimes.
We checked that replacing our default phase tracing choice by it does not impact the presented results.
For $g \leq 0.6$ it reproduces $v$ to within $0.2\%$, $\Delta V$ to within $0.04\%$, and $\Treh$ to within $0.2\%$.}

The energy and entropy densities of each phase follow from $V$ at its respective minimum, together with the SM
plasma at the same
temperature, whose degrees of freedom we take from the tabulated lattice data of ref.~\cite{Saikawa:2018rcs}.
Some caution is required with this approach, as the high-temperature expansion is only valid in the symmetric
phase. Hence, the daisy term, which is derived in the high-temperature expansion, is invalid in the broken
phase at low temperature where $g\phi \gg T$.
There, the dark photon contribution to the last term in eq.~\eqref{eq:V4D} becomes proportional to $T^3$
at low $T$, leading to an unphysical scaling of the dark sector energy density with $T^3$ instead of $T^4$,
corresponding
to a speed of sound $\cs^2 = 1/2$ instead of $1/3$. We therefore omit the daisy term in the energy density, while
still including it in the phase tracing.\footnote{Including it in the energy density, or computing the thermal
masses from $\langle\phi^2\rangle_T(m)/\langle\phi^2\rangle_T(0) = (6/\pi^2)\sum_n
(x/n) K_1(nx)$ with $x = m/T$, which effectively suppresses each contribution by a Boltzmann
factor at low $T$~\cite{Curtin:2016urg, Ringwald:2020vei}, changes $\Treh$ by at
most $0.3\%$, and $\Tp$ and $\bHR$ by less than $10^{-4}$. Within the Parwani scheme~\cite{Parwani:1991gq},
which resums the thermal masses into every $m_i^2$ of the one-loop terms and thereby avoids the term $\propto T^3$,
$\Tp$, $\Treh$ and $\bHR$ change by less than a relative $10^{-4}$ at the NANOGrav best fit.}

\subsection{Bubble nucleation rate}
\label{sec:rate}

A first-order phase transition proceeds through the nucleation of bubbles.
The bubble nucleation rate per unit volume and time, $\Gamma(T)$, sets the number
of bubbles surpassing the size of the critical bubble, at
a temperature $T$ of the plasma outside them. For a thermal phase transition it
is given by~\cite{Linde:1980tt,
Linde:1981zj, Affleck:1980ac, Langer:1969bc, Ekstedt:2021kyx, Gould:2021ccf}
\begin{equation}
  \Gamma = \Adyn\,\Astat\,\eu^{-\Sthree/T}\,,
  \label{eq:rate}
\end{equation}
where $\Sthree$ is the action of the critical bubble, i.e.~the bounce.
The statistical prefactor $\Astat$ follows from the ratio of the functional
determinants of the fluctuation operator around the critical bubble and around the false vacuum, with
the zero modes replaced by integrals over their collective coordinates. Herein, the three zero modes that shift the
bubble in space are replaced by an integral over its position, which results in a volume factor.
The dynamical prefactor $\Adyn$
describes the real-time dynamics.
It accounts for the rate at which a critical bubble, once formed, grows away from the saddle point
instead of shrinking back.

The three-dimensional EFT describes the soft modes of the plasma, with momenta of order $gT$, i.e.~the
length scales at which nucleation takes place, while modes of smaller length scales are integrated out. Its
parameters follow from those of the four-dimensional theory by matching at the scale $\mu_4 = 4\pi
\eu^{-\gamma_{\rm E}}\,T \simeq 7.06\,T$, with $\gamma_{\rm E}$ the Euler--Mascheroni constant, and its
own renormalisation scale being $\mu_3 = gT$.
We distinguish two orders in this work.
At leading order (LO), the bounce is
computed with the LO matching relations and the one-loop soft potential $V_3$ of eq.~\eqref{eq:V3}, in
which the gauge field is integrated out, and the prefactors are estimated by dimensional analysis. At
next-to-leading order (NLO), the matching relations include their one-loop corrections, and the rate
contains the functional determinant of the fluctuation operator around the bubble, the two-loop
potential, a subtraction $C$ that removes from the determinant the contributions already contained in
$V_3$, and the sextic operators with coefficients $c_6$ and $c_D$ of the soft theory. By default we compute the nucleation rate with NLO precision~\cite{Ekstedt:2021kyx,
Kierkla:2025qyz, Puchades-Ibanez:2026ksh}, with the functional determinants from the Gelfand--Yaglom
method~\cite{Gelfand:1959nq}.

At either order the rate is written in the form
\begin{equation}
  \Gamma = T^4\left(\frac{S_{\rm tot}}{2\pi}\right)^{3/2}\eu^{-S_{\rm tot}}\,,
  \label{eq:Stot}
\end{equation}
which is the input of the percolation calculation discussed in the following section. The prefactor is the
Jacobian of the three translational zero modes times the dimensional estimate $T^4$ for the remaining
factors~\cite{Linde:1980tt, Linde:1981zj}. At LO the exponent is expressed as $S_{\rm tot} = S_{\rm LO} \equiv
\Sthree^{\rm LO}/T$, with $\Sthree^{\rm LO}$ the action of the bounce of the LO-matched theory. At NLO it is
$S_{\rm tot} = S_{\rm NLO} \equiv \Sthree^{\rm NLO}/T + \Delta S$, with $\Sthree^{\rm NLO}$ the action of
the bounce of the NLO-matched theory and $\Delta S$ the correction from the functional determinant, the
two-loop potential, the subtraction $C$, the operators of dimension six, and the prefactors $\Adyn$ and
$\Astat$, defined by equating eq.~\eqref{eq:Stot} with
eq.~\eqref{eq:rate}. For the
dynamical prefactor we take the functional form derived by Langer~\cite{Langer:1969bc, Affleck:1980ac,
Hirvonen:2025hqn},
$\Adyn = \sqrt{|\lambda_-|}/2\pi$, with $\lambda_-$ the negative eigenvalue of the fluctuation operator
around the critical bubble, which approximates to $\Adyn \simeq 0.1\,T$ throughout our parameter space.

\begin{figure}[t]
\centering
\includegraphics[width=\textwidth]{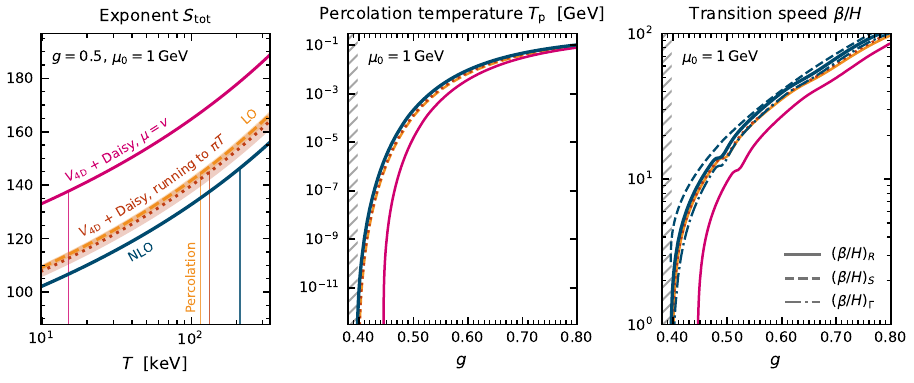}
\caption{\emph{Left}: The exponent $S_{\rm tot}$ of the nucleation rate, eq.~\eqref{eq:Stot}, at $g = 0.5$ and
$\muz = 1$\,GeV against the temperature, at NLO (solid), at LO (dashed), and with $V_{\rm 4D}$ + Daisy (see
eq.~\eqref{eq:V4D}), with the couplings run at one loop to $\mu = \pi T$ (\lk{N4}, dotted) and renormalised
at the vacuum scale $\mu = v$ (\lk{N5}, solid). Bands: the matching scale $\mu_4$ halved and doubled for LO and NLO,
the scale $\pi T$ halved and doubled for \lk{N4}. No band is shown for \lk{N5}, see the main text for details.
The percolation temperature is indicated by a vertical line in each case. \emph{Middle} and \emph{right}: 
The percolation temperature and the speed $\bHR$ for $\muz = 1$\,GeV from the mean bubble separation, in the
same colours as in the left panel. The right panel also shows the
speeds $\bHS$ (dashed) and $\bHG$ (dash-dotted) of eq.~\eqref{eq:speeds}
of the same NLO nucleation rate.
The hatched areas at $g \lesssim 0.4$ indicate that the phase transition with the NLO rate does not
complete, and the vacuum stays trapped.
\label{fig:exponent}}
\end{figure}

The left panel of figure~\ref{fig:exponent} shows both exponents at $g = 0.5$, together with those of the two
rates based on the 4D potential in variants \lk{N4} and \lk{N5} compared in detail in section~\ref{sec:alt_rate}.
The NLO action $S_{\rm NLO}$ runs nearly parallel to $S_{\rm LO}$, below it by $|\delta
S| = |S_{\rm NLO} - S_{\rm LO}| = 7$ to $10$ at the percolation temperature $\Tp$ (defined in
section~\ref{sec:perc}) for $g$ between $0.44$ and $0.5$.
$\Delta S$ contributes about one unit to the shown difference $\delta S$.
The exponent of $V_{\rm 4D}$ + Daisy with the couplings run to $\pi T$ (\lk{N4}) lies between
the LO and the NLO exponents, and that with the couplings fixed at the vacuum scale (\lk{N5}) about $30$ units above
the NLO exponent. At NLO the exponent depends quadratically on $\ln\mu_4$, with its minimum near $1.4\,\mu_4$, so
that halving and doubling $\mu_4$ both raise it. The NLO curve is the lower edge of its band, which is $0.12$ wide in
the exponent at $\Tp$ and hence practically invisible.
For \lk{N5} no band is shown.
With the quartic coupling fixed by the condition that the minimum lies at $\phi = v$, halving or doubling the
renormalisation
scale $\mu$ is absorbed into $\lambda$ up to the thermal terms,
and the band would be narrower than the plotted line.
Appendix~\ref{app:rate} summarises the computation of the nucleation rates in the EFT and compares them
with the independent computations of refs.~\cite{Christiansen:2025xhv, Puchades-Ibanez:2026ksh}. The
remaining panels of the figure are discussed in the next subsection.

\subsection{Percolation, reheating, and bubble separations}
\label{sec:perc}

The nucleation rate determines the three key quantities
that enter the GW spectrum:
\textit{i)} the percolation temperature $\Tp$, at which the bubbles form a connected cluster,
\textit{ii)} the
speed $\bH$ of the phase transition, its inverse duration in units of the Hubble time, and
\textit{iii)} the reheating temperature
$\Treh$, the temperature of the plasma once the vacuum energy has been converted into radiation. All three are based on
the knowledge of the true-vacuum fraction $P_{\rm t}(t) = 1 - \eu^{-I(t)}$, i.e.~the probability of a given
point in space
to already have converted to the true vacuum at the time $t$. For bubbles expanding at the speed of light,
\begin{equation}
  I(t) = \frac{4\pi}{3}\int^t_0 \dd t'\,\Gamma(t')\,a(t')^3\left[\eta(t) - \eta(t')\right]^3
  \label{eq:percInt}
\end{equation}
is the volume of all bubbles per unit volume in the past lightcone, counting overlapping regions
repeatedly~\cite{Guth:1981uk, Ellis:2018mja}. Here,
$a$ is the scale factor and $\eta$ the conformal time, so that $\eta(t) - \eta(t')$ is the comoving radius
at the time $t$ of a bubble nucleated at $t'$. The Hubble rate follows from the spatially averaged energy density of
the two phases~\cite{Matuszak:2026xsz},
\begin{equation}
  3M_{\rm Pl}^2H^2 = (1 - P_{\rm t})\,\rho_{\rm f} + P_{\rm t}\,\rho_{\rm t}\,,
  \label{eq:friedmann}
\end{equation}
with $M_{\rm Pl} = 2.4\times10^{18}$\,GeV the reduced Planck mass. Here,
$\rho_{\rm f}$ is the energy density outside the bubbles and
it contains the vacuum energy of the false minimum and the plasma radiation energy density. The
energy density $\rho_{\rm t}$ within the bubbles
is fixed by the plasma of the converted regions, into which the vacuum energy enters as
radiation at the moment of conversion and then redshifts. \TL\ integrates the
true-vacuum fraction self-consistently on this background. To express $\Gamma(T)$ as a
function of time, temperature and time are related via entropy
conservation, $a \propto 1/(T h_{\rm
eff}^{1/3})$ with $h_{\rm eff}$ the entropy degrees of freedom, for a
dark sector in kinetic equilibrium
with the SM plasma.

The phase transition percolates once $P_{\rm t}$ reaches the
percolation threshold $\fperc \simeq
0.29$~\cite{Rintoul:1997tze}, at the percolation temperature $\Tp$,
and completes once $P_{\rm t}$ reaches $0.99$,
at the final temperature $T_{\rm f}$. For transitions which do not
reach the completion threshold, the vacuum state is referred to as
being trapped in the false state.

The middle panel of figure~\ref{fig:exponent} shows $\Tp$ as a
function of $g$
for a fixed $\muz = 1$\,GeV, i.e.~for $v = \eu^{1/6}\,{\rm GeV}/g$.
The percolation temperature drops quickly
towards the left border of the panel, from $\Tp \simeq 0.1$\,GeV
at $g = 0.8$ to about $7 \, \text{meV}$ at
$g = 0.40$ for the NLO rate. 
Since the NLO exponent is smaller than the LO one, it predicts percolation
at a slightly higher temperature. The other rates follow the same pattern.
Below $g \simeq 0.4$, the NLO nucleation rate predicts that the transition
does not complete, and the vacuum remains in a trapped state. The other
rates predict completion thresholds at slightly larger values of $g$, as
shown in the figure.
Most notable is the large difference between the percolation temperatures
of the NLO and \lk{N5} (4D + Daisy at $\mu = v$) rates, which is due to the large difference in
their exponents, as shown in the left panel of figure~\ref{fig:exponent}.
With that nucleation rate, the completion threshold lies at $g \simeq 0.45$, i.e.~roughly 10\% higher
than for the NLO rate. This shift will appear again in the preferred regions
of the parameter space in section~\ref{sec:preferred} and in the discussion
of variant \lk{N5} in section~\ref{sec:alt_rate}.

The strength of the phase transition, approximated by
$\alpha \simeq \Delta V/\rho_{\rm rad}(\Tp)$, is the ratio of the released vacuum
energy to the energy density of the radiation in the false vacuum at $\Tp$. A phase transition
with $\alpha \gg 1$ is referred to as being strongly supercooled. In the present case
we exclusively find strong supercooling, $\alpha > 10^6$ within the preferred regions, up to values
around $\alpha \simeq 5\times10^{36}$. Hence,
$\alpha/(1+\alpha) \rightarrow 1$ for every definition of $\alpha$ in the literature~\cite{Giese:2020rtr,
Giese:2020znk} and the kinetic energy fraction saturates. Therefore, the GW spectrum
does not depend on $\alpha$ in this model~\cite{Ellis:2019oqb, Lewicki:2022pdb} and
special care has to be taken when choosing the relevant GW spectral template, as described in the
following subsection.

The speed $\bH$ denotes the inverse duration of the phase transition in units of the Hubble
time $H^{-1}$ at percolation. For a nucleation
rate that grows \emph{exponentially} in time, $\Gamma \propto \eu^{\beta t}$, $\beta$ is its growth rate.
For nucleation rates that are
\emph{not} exponential in time and for slow phase transitions ($\bH \lesssim 100$) there are several
definitions of the speed to relate the speed to the mean bubble separation $R_\star$, which
sets the peak frequency of the GW signal. With $\nb$ the number density of
bubbles nucleated until $\Tp$, so that $R_\star = \nb^{-1/3}$, we use
\begin{equation}
  \bHR = \left(\frac{8\pi}{\fperc}\right)^{1/3}\frac{1}{R_\star H}\,,\qquad
  \bHS = T\,\frac{\dd S_{\rm tot}}{\dd T}\bigg|_{\Tp}\,,\qquad
  \bHG = \frac{1}{H}\,\frac{\dd\ln\Gamma}{\dd t}\bigg|_{\Tp}\,.
  \label{eq:speeds}
\end{equation}
The first definition $\bHR$ is based on the mean bubble separation, the second $\bHS$
on the slope of the exponent of the nucleation rate, and the third $\bHG$ on the slope
of the nucleation rate itself. The three definitions coincide for a large exponential
nucleation rate, but differ for a small non-exponential one. The latter is the generic case
when studying observable GW signals~\cite{Matuszak:2026xsz}. In particular, $\bHR$ is always
positive, while $\bHS$ and $\bHG$ can be negative if the nucleation rate decreases at
the percolation temperature~\cite{Matuszak:2026xsz}, which in the present model happens for
$\bHG$ in a narrow strip along the edge of vacuum trapping at $g \lesssim 0.42$.

The right panel of figure~\ref{fig:exponent} compares
these definitions along $g$ for the benchmark line with $\muz = 1$\,GeV.
All three quantities drop towards the left boundary of the plot, which is 
limited by the vacuum-trapped region of parameter space. Moreover, we find 
$\bHS > \bHR > \bHG$ at every coupling. At $g = 0.40$, the
smallest coupling with a completing transition resolved in this plot,
the NLO rate gives $\bHS = 4.4$, $\bHR = 1.8$ and $\bHG = 0.36$. For
$g \geq 0.45$ the LO rate and the rate of \lk{N4} (4D + Daisy, $\mu = \pi T$)
predict speeds $\bHR$ about $10\%$ below the NLO one. As for the
percolation temperature, a larger exponent delays percolation to a
lower temperature, where $S_{\rm tot}$ varies more slowly with $T$, so that the
phase transition proceeds more slowly. The \lk{N5} (4D + Daisy, $\mu = v$) prediction of the
transition speed $\bHR$ is again roughly $10\%$ shifted towards larger gauge couplings, compared
to the NLO curve. The features in the NLO $\bHR$ between $g = 0.48$ and $0.52$
and between $g = 0.61$ and $0.70$ are due to the degrees of freedom of the SM
plasma dropping at the electron-positron annihilation and at the
annihilation of the muons, respectively. Since $\Tp \lesssim 100 \, \text{MeV}$ for
all points on this line, the QCD crossover only affects the largest shown
$g \gtrsim 0.75$. In either of these cases, the entropy released into the plasma
slows the decrease of the temperature in time and thereby lowers $\bHR$ slightly.

The last quantity read off at percolation is the reheating temperature, which sets
the redshift of the GW signal. \TL\ follows the temperature
$T_{\rm b}$ of the converted
regions through the conversion of the vacuum energy. With $T$ the temperature of
the false-vacuum plasma, which
falls as $1/(a h_{\rm eff}^{1/3})$, with the entropy degrees of freedom of the
false-vacuum plasma, it integrates
\begin{equation}
  \frac{\dd T_{\rm b}}{\dd T} = \frac{\dd\ln s_{\rm s}}{\dd T}\,\frac{s_{\rm b}(T_{\rm b})}{s_{\rm b}'(T_{\rm b})}
  + \frac{\dd\ln P_{\rm t}}{\dd T}\,\frac{\rho_{\rm s}(T) - \rho_{\rm b}(T_{\rm b})}{\rho_{\rm b}'(T_{\rm b})}\,,
  \label{eq:Tb}
\end{equation}
with $s_{\rm b}$ and $\rho_{\rm b}$ the entropy and energy density of the
broken phase, $s_{\rm s}$ and $\rho_{\rm s}$
those of the false vacuum, including the vacuum energy, and primes denoting
derivatives with respect to $T_{\rm b}$. The first
term redshifts the radiation of the converted regions at constant entropy,
and the second adds the rate of liberated vacuum energy. The integration
starts at $P_{\rm t} = 0$, with $\rho_{\rm b}(T_{\rm b}) = \rho_{\rm s}(T)$. The
reheating temperature is then taken as $\Treh = T_{\rm b}(\Tp)$,
the temperature within the bubbles at percolation.

\subsection{Gravitational wave spectrum}
\label{sec:source}

The GW spectrum is predicted by mapping the relevant time, length, and energy scales of the phase transition to
semi-analytical templates based on lattice simulations. Every such template maps the
nucleation history $\Gamma(T)$ and the thermodynamics of the phase transition, compressed into a few
parameters read off at
one moment ($\alpha$, $\bHR$, $\Treh$, and efficiency factors $\kappa$ at $\Tp$), to a GW spectrum.
For the phase transition of
the present conformal model this mapping is not guaranteed to hold.
Many of the available templates are only valid within specific
limits of these parameters, typically $\beta/H \gg 100$ and $\alpha \ll 1$,
and the simulations are usually initialised with either 
an exponentially growing bubble nucleation rate, or the simultaneous nucleation of
bubbles~\cite{Hindmarsh:2013xza, Hindmarsh:2015qta, Hindmarsh:2017gnf, Cutting:2019zws, Jinno:2022mie,
Caprini:2024gyk, Caprini:2009yp, RoperPol:2019wvy, Auclair:2022jod}.
Neither of these assumptions
holds in the present model.
Instead $\alpha \gg 1$, $\bHR \lesssim 100$ and $\Gamma(T)$ is non-exponential, see figure~\ref{fig:exponent}.
Hence, the phase transition is so strongly supercooled and completes over such a long timescale that the
expansion of the
universe during the phase transition matters and different spectral templates have to be employed.

The closest template to matching these requirements which is available today is that of ref.~\cite{Lewicki:2025hxg},
which is based on simulations of a strongly supercooled phase transition in vacuum, with $\alpha \gg 1$
and $\bHR \gtrsim \mathcal{O}(5)$, capturing the effect of Hubble expansion during the phase transition.
These templates can be understood as a direct extension of previous work~\cite{Lewicki:2022pdb},
based on the bulk flow model~\cite{Konstandin:2017sat, Jinno:2017fby}, which indicated
that the GW spectrum from bubble walls and fluid shells is indistinguishable in the
limit of luminal wall velocities with large Lorentz boost factors.

In the present case of the conformal $U(1)$ model, we find that for $g \gtrsim 0.45$--$0.5$
(the precise value depending on $v$) the
bubble walls reach a terminal velocity before they collide, due to the friction
from dark gauge bosons radiated by particles crossing the
walls~\cite{Bodeker:2017cim, Ellis:2019oqb, Azatov:2023xem}. There, the walls carry only
a small fraction of the released energy when they collide, and the remaining energy is
carried by thin relativistic shells of dark gauge bosons radiated by the 
walls~\cite{Gouttenoire:2021kjv}. Hence, in this case the GW spectrum is dominated by the
collision of these shells, which is captured by the template of ref.~\cite{Lewicki:2025hxg}.
For $g \lesssim 0.45$, the walls collide before reaching their terminal velocity, and they
carry most of the released energy. In this case, the GW spectrum is dominated by the collision
of the bubble walls themselves, which is also captured by the same template.
We hence treat the efficiency of the GW source as $\kappa = 1$, as the GW signal
does not distinguish between the two cases~\cite{Lewicki:2022pdb}. We further treat
the energy-momentum tensor of the source as falling with the radius of the bubble as
$R^{-3}$, as in the dissipative bulk-flow model~\cite{Lewicki:2022pdb}, which is appropriate
for the present case of a conformal $U(1)$ model with relatively large gauge couplings, as
shown in refs.~\cite{Lewicki:2020azd,Baldes:2024wuz}.

We therefore by default employ the fitted template of ref.~\cite{Lewicki:2025hxg} for the
dissipative bulk flow model including the effect of Hubble expansion within
our subsequent analysis. The GW spectrum is given by a broken power law
\begin{equation}
  h^2\OGW(f) = \frac{\mathcal{R}h^2\, A}{(\bHR)^2}
  \frac{(n_{\rm l}+n_{\rm h})^w}{\left[n_{\rm h}\,(f/f_{\rm p})^{-n_{\rm l}/w} + n_{\rm l}\,(f/f_{\rm p})^{n_{\rm h}/w}\right]^w}
  \left[1 + \left(\frac{f_{\rm H}}{f}\right)^3\right]^{-\frac{3-n_{\rm c}}{3}},
  \label{eq:bpl}
\end{equation}
with $\OGW$ the GW energy density per logarithmic frequency interval in units of the critical density,
$h$ the Hubble constant in units of $100$\,km\,s$^{-1}$\,Mpc$^{-1}$, $f_{\rm p}$ the peak frequency,
$n_{\rm l} = n_{\rm h} = w = 2$ the slopes below and above the peak and the width of the peak, and $n_{\rm
c}$ the logarithmic slope of the broken power law at the horizon frequency $f_{\rm H}$. 
Below $f_{\rm H}$ the spectrum
scales as
$f^3$.\footnote{
The causal tail is also impacted by the equation of state parameter when the
modes enter into the horizon after the phase
transition~\cite{Hook:2020phx, Franciolini:2023wjm}, which we do not take into
account in this work.
We checked that replacing the causal $f^3$ tail by a continuation of the slope $f^1$ or $f^2$
hardly moves the preferred NANOGrav region, since $f_{\rm H} \simeq 3$\,nHz,
just above the first NANOGrav bin. The IPTA region is not affected at all.
}
The factor
$\mathcal{R}h^2$ parametrises the redshift of the energy density from reheating to today, and the frequencies scale with
$a_{\rm p}H_{\rm p}/a_0$, the comoving Hubble rate at percolation over the scale factor today, which \TL\
evaluates for radiation domination starting at $\Treh$~\cite{Caprini:2019egz},
\begin{equation}
\begin{aligned}
  \mathcal{R}h^2 &= \Omega_\gamma h^2\left(\frac{h_{\rm eff,0}}{h_{\rm eff,reh}}\right)^{4/3}
  \left(\frac{g_{\rm eff,reh}}{g_{\rm eff,0}}\right),\\
  \frac{a_{\rm p}H_{\rm p}}{a_0} &= 1.65\times10^{-5}\,{\rm Hz}\left(\frac{\Treh}{100\,{\rm GeV}}\right)
  \left(\frac{g_{\rm eff,reh}}{100}\right)^{1/2}\left(\frac{100}{h_{\rm eff,reh}}\right)^{1/3} \, .
\end{aligned}
  \label{eq:redshift}
\end{equation}
Here, $\Omega_\gamma h^2$ is the photon energy density today in units of the critical density and the indices
reh and $0$ denote the degrees of freedom at reheating and today, respectively. The amplitude, the peak
frequency, and the horizon frequency are given by~\cite{Lewicki:2025hxg}
\begin{equation}
\begin{gathered}
  A = 0.06\left[1 + 0.8\left(1 - \eu^{\bHG^{-1/2}}\right)\right],\\
  \frac{2\pi f_{\rm p}}{a_{\rm p}H_{\rm p}} = 0.7\,\bHR\left[1 + 1.8\,\bHG^{-1.2}\right],\qquad \text{and} \qquad
  \frac{2\pi f_{\rm H}}{a_{\rm p}H_{\rm p}} = 1 - 0.6\,\bHG^{-0.67}.
\end{gathered}
\label{eq:lv}
\end{equation}
These semi-analytical fits are based on simulations with an exponential nucleation rate
$\Gamma(t) = H_\mathrm{V}^4 \eu^{\beta t}$
on a background expanding with a constant Hubble rate $\HV$ as in vacuum domination.
The transition speed hence enters in
two roles, which we keep separate. In the prefactor $A/\bHR^2$ of eq.~\eqref{eq:bpl} and in
the factor $0.7\,\bHR$ of $f_{\rm p}$ in eq.~\eqref{eq:lv} it stands for the mean bubble separation,
as it is set through the length scale $R_\star$ which sets the peak frequency of the spectrum. We
therefore use $\bHR$ in both places. In the square brackets of
eq.~\eqref{eq:lv}, instead, which describe the effect of the expansion of the universe during the
phase transition and which do not appear in the fit of the previous work~\cite{Lewicki:2022pdb}
ignoring the expansion, the speed stands for the growth rate of the nucleation rate, which sets the
time over which bubbles can nucleate, and
there we pass $\bHG$. 
The bracket
of $A$ in eq.~\eqref{eq:lv} is roughly $0.5$ at $\bHG = 4$ and becomes negative at
$\bHG \simeq 1.5$, below which \TL\ sets the amplitude to zero. 
The fits are written in $\beta/\HV$, with $\HV = (\Delta V/3)^{1/2}/M_{\rm Pl}$ the Hubble rate
of the vacuum energy alone, whereas $\bHG$ is normalised to the Hubble rate $H_{\rm p}$ of the mean
expansion at percolation, so that $\beta/\HV = \bHG\,H_{\rm p}/\HV$. In the $2\sigma$ regions $H_{\rm p}/\HV$
lies between $0.90$ and $0.98$, since part of the vacuum energy has already been released and redshifted at percolation.
The simulations cover $\beta \geq 4\HV$, and slower phase transitions are
included by extrapolating the template.

\subsection{Pulsar timing data and binary background}
\label{sec:data}

The last step of the analysis compares the predicted spectrum with the GW spectra inferred from pulsar timing data.
A free
spectrum is the posterior of the GW amplitude in each frequency bin after the pulsar and interstellar medium
noise has been
modelled and marginalised over.
The more precise alternative, a joint analysis of the pulsar timing data with
the red noise parameters of every pulsar kept free, as in the code \texttt{enterprise} used by
NANOGrav~\cite{NANOGrav:2023icp}, is
impractical as more than a hundred red noise parameters would have to be fit simultaneously.
We hence employ the likelihood based on the free spectra 
shipped with the codes \texttt{binaryfog}~\cite{Bringmann:2026uxo} and \texttt{ceffyl}~\cite{Lamb:2023jls}.
More information on the likelihoods is provided in appendix~\ref{app:preference}.

By default we fit the first $14$ Fourier frequency bins of the NANOGrav 15\,yr data set~\cite{NANOGrav:2023hde,
NANOGrav:2023gor} and the first $13$ of the second data release of the IPTA (IPTA DR2)~\cite{Perera:2019sca,
Antoniadis:2022pcn}, the bins to which the PTA analyses restrict their searches since above these the inferred
signal is contaminated by pulsar white noise~\cite{Antoniadis:2022pcn,Mitridate:2023oar}.
For both the NANOGrav and IPTA data sets we use the most recent release of their free spectra, of
2026~\cite{NG15KDE2, NANOGrav:2026err} and 2022, respectively, through the
code \texttt{binaryfog}~\cite{Bringmann:2026uxo}.
For
both we use the free spectra of the common uncorrelated red noise (CURN), since it carries the most spectral
information among the available PTA likelihoods~\cite{NANOGrav:2023gor, NANOGrav:2023hde}.

The binary contribution to the total signal is described through the population model emulated in
\texttt{binaryfog}~\cite{Bringmann:2026uxo}, based on the SMBHB population synthesis tool \texttt{holodeck}
developed by the NANOGrav collaboration~\cite{holodeck}.
We use the parametrisation of the binary signal of refs.~\cite{NANOGrav:2023hfp, NANOGrav:2026err}.
In the notation of ref.~\cite{Bringmann:2026uxo} these are the normalisation $\psi_0$ and the
characteristic mass $m_{\psi,0}$ of the galaxy stellar mass function (which count the galaxy mergers), the
normalisation $\mu$ and the intrinsic scatter $\epsilon$ of the relation between the black hole mass and the
bulge mass (which set the masses of the black holes) and the hardening timescale $\tau_{\rm f}$ and the index
$\nu_{\rm inner}$ of the binary hardening at small separations (setting the efficiency of the hardening process).
The first four can be inferred from galaxy surveys and mostly set the amplitude of the
background, while the last two are phenomenological and affect both the binary background's amplitude and slope at
low frequencies.
Typically, the first four parameters are therefore modelled through Gaussian priors based on measured survey
data, whereas the phenomenological hardening parameters have uniform priors.
The range of the latter, however, determines the flexibility of the binary background to fit the data and
hence requires a dedicated discussion which we will provide in section~\ref{sec:alt_binaries}.

Following the approach described in ref.~\cite{Bringmann:2026uxo} we use three variants of astrophysical prior
sets in this work
(see table~\ref{tab:binpriors} in the appendix).
The galaxy survey-averaged priors include the most recent measurements of the galaxy mass function and the
black hole--bulge mass relation.
The alternative priors of NANOGrav's own analysis~\cite{NANOGrav:2023hfp} have widths about
four times those of the fits they are centred on. The Tomczak--Kormendy priors are Gaussians at the centres and widths
of specifically these fits without any additional widening~\cite{Tomczak:2013bxa,Kormendy:2013dxa}.
The survey-averaged prior set gives a larger share of the signal to the binaries and hence represents a
slightly more optimistic view on the binary background than the NANOGrav priors.
For further details on these priors we refer to the appendix of ref.~\cite{Bringmann:2026uxo}.

Since the binary contribution strongly affects the parameter inference, our default fit contains the phase
transition alone,
so that in section~\ref{sec:alternatives} we can separate the effect of the other modelling choices.
The binaries enter as alternatives there.

\begin{figure}[t]
\centering
\includegraphics[width=\textwidth]{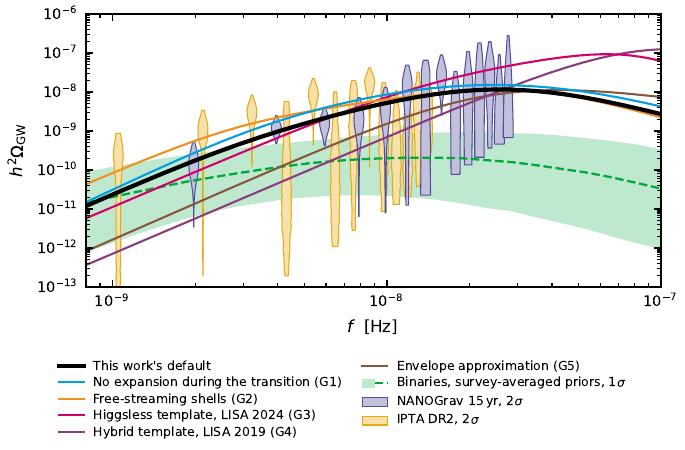}
\caption{Gravitational wave spectrum $h^2\OGW$ at the NANOGrav best fit ($g = 0.45$, $v =
2.9$\,GeV, $\bHR = 9.6$, $\Treh = 150$\,MeV) with this work's default template (black, eqs.~\eqref{eq:bpl}
and~\eqref{eq:lv}) and the alternative templates of section~\ref{sec:alt_source}: without expansion
during the phase transition (\lk{G1}), with free-streaming shells (\lk{G2}), the envelope approximation
(\lk{G5}), and the higgsless sound-wave template of the analysis pipeline of LISA (\lk{G3}) and the hybrid one
of the 2019 LISA update (\lk{G4}).
Violins: Bayesian posteriors of the free spectral amplitudes inferred from the NANOGrav 15\,yr and the second
IPTA data release.
Green: the background of the binaries predicted under the survey-averaged priors of table~\ref{tab:binpriors},
median (dashed) and
$1\sigma$ band over the prior and over realisations of the population. \label{fig:spectra}}
\end{figure}

In order to get a more intuitive understanding of the relevance of the binary background on the fit, in
figure~\ref{fig:spectra} we show a comparison of GW spectra and the binary background prediction.
The different GW spectra used throughout this work are shown for a benchmark point with $g = 0.45$ and $v =
2.9$\,GeV, which is the best-fit point of the NANOGrav data under all default modelling choices.
The dissipative bulk-flow spectrum including the effect of expansion, as described in the previous subsection,
is shown in black, while alternative spectra are illustrated with different colours and narrower line widths.
The spread between the different spectra already indicates that the dominant effect on the parameter inference
is through the choice of the spectral template, as discussed in detail in
the alternatives \lk{G1}--\lk{G5} in section~\ref{sec:alt_source}.
The $1\sigma$ band of the expected contribution from binaries, illustrated by the green band, follows the
\texttt{binaryfog} prediction based on the survey-averaged prior.
The NANOGrav 15\,yr (purple) and IPTA DR2 (yellow) free spectral likelihoods are visualised through Bayesian
spectrograms, often referred to as ``violins'', depicting the posterior distribution of the GW amplitude at
integer multiples of the lowest frequency.

The GW spectrum of the phase transition is steeply rising in the nanohertz band, and the PTA data fix its
amplitude and slope.
The fourteen NANOGrav bins span the range from $2$ to $28$\,nHz and the inferred phase transition signal peaks
at around $26\, \text{nHz}$, such that the bins cover
the rising slope of the spectrum.
The median of the expected binary background lies about two orders of magnitude below the measured peak amplitude, its
$1\sigma$ band stretches about one order of magnitude in each direction and its upper edge still lies about
one order of magnitude below the inferred peak amplitude.
This observation already allows for a rough estimate of the significance in favour
of a phase transition contribution of about $2\sigma$, to be quantified more
precisely in section~\ref{sec:alt_binaries}.
The
alternative GW templates illustrated in figure~\ref{fig:spectra} will be discussed in
section~\ref{sec:alt_source}.

\subsection{Preferred regions with the default modelling choices}
\label{sec:preferred}

\begin{figure}[t]
\centering
\includegraphics[width=\textwidth]{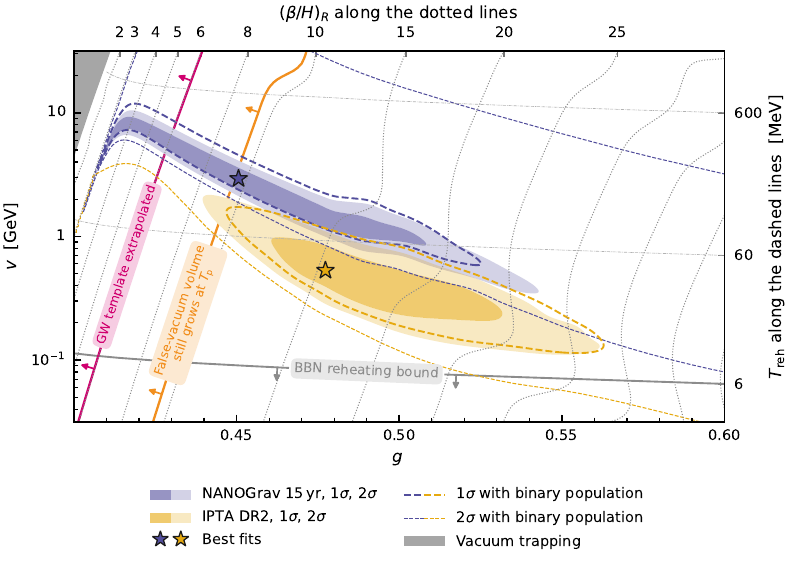}
\caption{The $1\sigma$ and $2\sigma$ regions of the $(g, v)$ plane preferred by the NANOGrav 15\,yr
(purple) and IPTA DR2 (yellow) data with every modelling choice at its default (filled contours), and with the binary
population of \texttt{binaryfog} marginalised over the survey-averaged priors (\lk{B1}, dashed: $1\sigma$
thick, $2\sigma$ thin).
With the binaries, the posterior keeps a constant value wherever the phase transition contributes little, up
to the edges of the prior range, $0.39 < g < 0.75$ and $10\,{\rm MeV} < v < 100\,{\rm GeV}$.
The $2\sigma$ contours thus enclose every point at which the fit is not worse than with the binaries alone.
Since the data do not exclude the binaries alone strongly enough, the $2\sigma$ contours depend on the prior
range, while the $1\sigma$ contours hardly do.
The profiled likelihood of appendix~\ref{app:profiling} avoids this dependence.
The best-fit points are indicated by stars.
Thin dotted and dashed lines in grey indicate the contours of equal $\bHR$ and $T_\mathrm{reh}$, as indicated
on the upper and right axes, respectively.
The thick solid lines in magenta mark the slowest simulated phase transition, $\bHR \simeq 6$, to the left of
which the template is extrapolated.
The analogous orange and grey lines show the parts of parameter space where the completion criterion of
ref.~\cite{Ellis:2018mja} fails, and where the BBN reheating bound is violated, respectively.
In the dark grey region the universe stays trapped in the false vacuum.  \label{fig:plane}}
\end{figure}

To conclude this section, here we present the regions of the $(g, v)$ plane that the NANOGrav and IPTA data
prefer with every modelling choice at its default, as described in the previous subsections.
Figure~\ref{fig:plane} shows the regions preferred by the two data sets.
The $1\sigma$ and $2\sigma$
regions are the regions of highest posterior density that contain $68\%$ and $95\%$ of the posterior mass
for the likelihood $L$ with flat priors over the ranges $0.39 < g < 0.75$ and $1 < \log_{10}(v / {\rm MeV}) < 5$.
For NANOGrav the marginalised $1\sigma$ intervals are $g = 0.43$ to $0.50$ and $v = 0.96$\,GeV to $4.8$\,GeV,
and the best fit lies at $g = 0.45$ and $v =
2.9$\,GeV, with $\bHR = 9.6$ and $\Treh = 150$\,MeV. In the main part of the $1\sigma$ region $\bHR$
ranges from $3.3$ to $18$, $\Treh$ from $62$\,MeV to $320$\,MeV and the dark photon mass $m_{A'}$ from $0.45$\,GeV to
$3.7$\,GeV.
Towards slower phase transitions the region extends beyond the slowest simulated phase transition, $\beta =
4\HV$, to the left of the magenta line in figure~\ref{fig:plane}, where the template
is extrapolated.
The narrow feature in the contour at small $g$, just below the vacuum trapped region, is due to $\bHG \simeq
1.5$, corresponding to $A \to 0$ (see eq.~\eqref{eq:lv}).
There, the decrease in the GW amplitude for smaller $g$ can be compensated by shifting the preferred region
towards lower frequencies, such that the spectrum's peak shifts towards lower frequencies without affecting
the goodness-of-fit in the low-frequency bins.
This prolongs the $v$ posterior downwards and results in the visible spike-like feature in the shown
contours.\footnote{With $\bHR$ instead of $\bHG$ in the expansion factors of eq.~\eqref{eq:lv}, the
zero-crossing of $A$ moves to $\bHR \simeq 1.5$, close to the vacuum trapped region, and the spike disappears.
The marginalised posteriors discussed in section~\ref{sec:alternatives} are hardly affected upon that change.
We note that also the formation of primordial black holes and the scalar-induced GW backgrounds may become relevant under
these conditions~\cite{Gouttenoire:2023naa, Gouttenoire:2023pxh, Franciolini:2025ztf, Wang:2026zvz,
Kierkla:2025vwp}.
Dedicated simulations for $\bHR \simeq \mathcal{O}(1)$ with the present non-exponential bubble nucleation rate
$\Gamma(T)$ would be required to resolve this region.}

Figure~\ref{fig:plane} also shows several restrictions on the preferred parameter space, none of which is
imposed by default in the fits of this work.
In the grey region in the upper left corner at small $g$, no final temperature
$T_{\rm f}$ exists as the true vacuum fraction never reaches the required threshold $P_\mathrm{t} = 0.99$.
There, the universe stays trapped in the false vacuum and no GWs are emitted (section~\ref{sec:perc}),
consistent with the used GW template including the effect of Hubble expansion during the transition.
The completion
criterion of ref.~\cite{Ellis:2018mja}, indicated in orange in the figure, is stricter.
It requires that the physical volume of the false vacuum,
proportional to $a^3(1 - P_{\rm t})$, already decreases at percolation. To the left of the orange line this volume
still grows at $\Tp$, although the phase transitions there eventually complete as $T_\mathrm{f}$ can be found
and the false vacuum fraction asymptotically reaches the required threshold (see also variant \lk{P3} of
section~\ref{sec:alt_perc}). 
The preferred regions moreover lie above the lower bound $\Treh > 6$\,MeV (grey line) from Big Bang
nucleosynthesis (BBN), the cosmic microwave
background and galaxy surveys~\cite{Barbieri:2025moq}, which only the $2\sigma$ region of IPTA approaches.
We checked explicitly that the BBN bound on $\Delta\Neff$ from the GW background itself does not constrain the
shown parameter space.
The largest $\Delta\Neff$ found on the plotted plane is $0.013$, more than an order of magnitude below the
relevant BBN and cosmic microwave background limits~\cite{Planck:2018vyg}.

Because the IPTA free spectrum prefers a lower peak frequency at around $10$\,nHz as well as a lower
amplitude, see figure~\ref{fig:spectra}, the posterior shifts towards faster phase transitions at
smaller $v$ with respect to the NANOGrav data set.
The IPTA best fit lies at $g = 0.48$ and $v = 0.53$\,GeV, and the $1\sigma$ intervals range from $g = 0.47$ to
$0.52$ and $v = 0.24$\,GeV to $0.70$\,GeV.
The speed $\bHR$ of the phase transition ranges from $12$ to $24$, $\Treh$ from $17$\,MeV to $62$\,MeV and $m_{A'}$
from $0.11$\,GeV to $0.46$\,GeV.
A joint fit of both data sets finds its best fit between the two, at $g = 0.48$ and $v = 1.1$\,GeV, where
$\ln L$ lies $2.1$ below the NANOGrav and $2.9$ below the IPTA maximum. For two shared parameters this corresponds
to a tension of $2.7\sigma$.
This value should only be understood as an indication since IPTA DR2 contains the NANOGrav 9\,yr data and is
therefore not statistically independent.
As discussed above this discrepancy is due to the difference in the inferred free spectra, possibly due to the
older data with fewer pulsars
and a different noise modelling.
The dashed contours of figure~\ref{fig:plane}, with the binaries included, are discussed together with variant
\lk{B1} in section~\ref{sec:alt_data}.

\section{Effect of alternative modelling choices on the parameter inference}
\label{sec:alternatives}

In this section we discuss by how much the inferred gauge coupling $g$ and vacuum expectation value $v$ depend on
the modelling choices adopted in section~\ref{sec:signal}.
In order to do so in a systematic way, we replace each of them, one at a time
as listed in table~\ref{tab:switches}, and repeat the fit.

As discussed in section~\ref{sec:preferred}, the PTA data fix mainly two properties of the spectrum, its peak
amplitude, set through $\bHR$, and its peak
frequency, set through the product $\bHR\,\Treh$. Moreover, some additional information is contained in the
low-frequency slope of the inferred GW spectrum, which helps in resolving the degeneracy between astrophysical and
cosmological signal sources. The alternative modelling choices presented in the following therefore move the
preferred region in the $(g, v)$ plane predominantly through their influence on the transition speed $\bHR$ and the
reheating temperature $\Treh$, and to a minor degree through their influence on the slope of the GW spectrum below
its peak.

\begin{figure}[t]
\centering
\includegraphics[width=\textwidth]{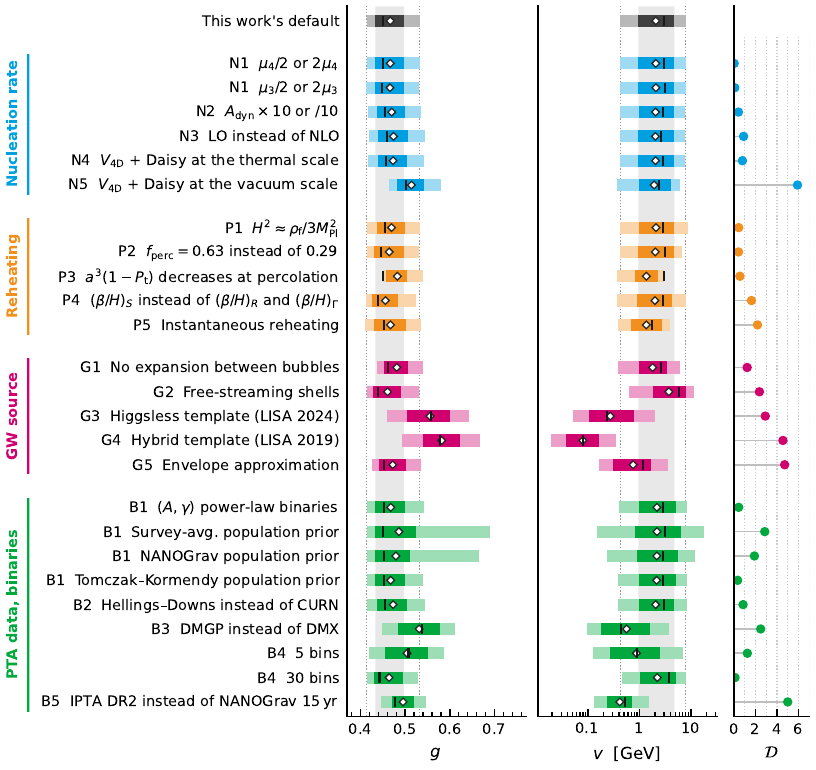}
\caption{Preferred regions of the NANOGrav 15\,yr data under each alternative of
table~\ref{tab:switches}: the marginal $1\sigma$ (dark) and $2\sigma$ (light)
intervals of $g$ (left)
and of $v$ (middle) against those of this work's default (grey band and dotted
lines), and the distance $\DM$
of eq.~\eqref{eq:distance} between the posterior means (right). Diamonds indicate the
posterior means, from which the distances $\DM$ are computed. Vertical ticks show the
values of $g$ and $v$ for the best fits within the given variant.
} \label{fig:budget}
\end{figure}

In order to describe the effect of the modelling choices, we quote not only the $1\sigma$ and $2\sigma$ intervals of
the alternative posterior distributions in $g$ and $v$, but also the Mahalanobis distance $\DM$ between the posterior
means. It is defined as
\begin{equation}
  \DM^2 = \Delta\boldsymbol\theta^{\rm T}\,\Sigma_0^{-1}\,\Delta\boldsymbol\theta\,,
  \label{eq:distance}
\end{equation}
with $\Delta\boldsymbol\theta$ the shift of the posterior mean of $\boldsymbol\theta = (g, \log_{10} v)$ and
$\Sigma_0$ the covariance of the default posterior. A shift $\DM = 1$ hence describes a
shift by one standard deviation of the posteriors depicted in figure~\ref{fig:plane}. Figure~\ref{fig:budget} shows
the posterior intervals and $\DM$ for every studied alternative modelling choice.
In table~\ref{tab:switch_numbers} in the appendix we provide the numerical values of the $1\sigma$ and
$2\sigma$ intervals and of $\DM$ for every alternative, the shift $\Delta \ln L$ in the goodness-of-fit to the
PTA which can be reached in that variant, as well as the ratio of areas of the $1\sigma$ regions to facilitate
the comparison with the default analysis of the previous section.
Sections~\ref{sec:alt_rate} to~\ref{sec:alt_data} follow the four steps depicted in
figure~\ref{fig:pipeline}, section~\ref{sec:alt_binaries} discusses the significance of a contribution of the
phase transition over the binaries alone, and section~\ref{sec:summary} summarises our findings and compares
them with earlier work.

\begin{table}[!t]
\centering
\footnotesize
\setlength{\tabcolsep}{1.8pt}
\begin{tabular}{@{}llll@{}}
\toprule
Label & Default & Alternative & Used in \\
\midrule
\alt{N}{1} & $\mu_4 = 7.06\,T$, $\mu_3 = gT$ & $\mu_4/2$, $2\mu_4$; $\mu_3/2$, $2\mu_3$ & \cite{Croon:2020cgk, Christiansen:2025xhv, Kierkla:2026bnm} \\
\alt{N}{2} & Langer prefactor & Ten times larger or smaller & -- \\
\alt{N}{3} & NLO & LO & (\cite{Kierkla:2023von, Puchades-Ibanez:2026ksh}) \\
\alt{N}{4} & 3D EFT & $V_{\rm 4D}$ + Daisy, couplings run to $\pi T$ & \cite{Christiansen:2025xhv} \\
\alt{N}{5} & 3D EFT & $V_{\rm 4D}$ + Daisy, renormalised at $\mu = v$ & \cite{Balan:2025uke, Goncalves:2025uwh, Bringmann:2026xcx} \\
\midrule
\alt{P}{1} & $\bar{\rho} = P_\mathrm{t} \rho_\mathrm{t} + (1 - P_\mathrm{t}) \rho_\mathrm{f} $ & $\bar{\rho} \simeq \rho_\mathrm{f} $, ignoring redshift of $\rho$ within bubbles & \cite{Kierkla:2022odc, Athron:2023mer, Puchades-Ibanez:2026ksh} \\
\alt{P}{2} & $\fperc = 0.29$ & $\fperc = 1 - 1/\eu \simeq 0.63$ & \cite{Lewicki:2024sfw, Lewicki:2025hxg, Kierkla:2026bnm} \\
\alt{P}{3} & $T_{\rm f}$ exists & $a^3(1 - P_{\rm t})$ decreases already at percolation~\cite{Ellis:2018mja} & \cite{Kierkla:2022odc, Athron:2023mer, Goncalves:2025uwh} \\
\alt{P}{4} & $\bHR$, $\bHG$ & $\bHS$ & \cite{Madge:2023dxc, Basler:2024aaf, Biondini:2026uds} \\

\alt{P}{5} & $\Treh$ of eq.~\eqref{eq:Tb} & Instantaneous reheating, $\Treh \simeq \Tp(1+\alpha)^{1/4}$ & \cite{Madge:2023dxc, Basler:2024aaf, Costa:2025csj, Biondini:2026uds} \\
\midrule
\alt{G}{1} & Expanding background & No Hubble expansion during the phase transition & \cite{Ellis:2023oxs, Lewicki:2024ghw, Puchades-Ibanez:2026ksh} \\
\alt{G}{2} & Dissipative shells & Free-streaming shells (bulk flow) & \cite{Gouttenoire:2023bqy, Christiansen:2025xhv} \\
\alt{G}{3} & Relativistic shells & Sound waves, higgsless simulations, LISA 2024 & \cite{Balan:2025uke, Goncalves:2025uwh} \\
\alt{G}{4} & Relativistic shells & Sound waves, hybrid simulations, LISA 2019 & \cite{Bringmann:2023opz} \\
\alt{G}{5} & Dissipative bulk flow & Envelope approximation~\cite{Lewicki:2022pdb} & \cite{Madge:2023dxc, Puchades-Ibanez:2026ksh} \\
\midrule
\alt{B}{1} & No binaries & $(A_{\rm bhb}, \gamma_{\rm bhb})$ power law or \texttt{binaryfog} population & \cite{NANOGrav:2023hvm, Balan:2025uke, Bringmann:2026xcx, Bringmann:2026uxo} \\
\alt{B}{2} & CURN & Hellings--Downs & \cite{NANOGrav:2023hfp, Ellis:2023oxs, Puchades-Ibanez:2026ksh} \\
\alt{B}{3} & DMX & Dispersion measure as a Gaussian process & (\cite{NANOGrav:2023gor, NANOGrav:2023hfp}) \\
\alt{B}{4} & $14$ bins (IPTA $13$) & $5$ or $30$ bins & \cite{NANOGrav:2023hfp, Bian:2023dnv}; (\cite{Madge:2023dxc, Conaci:2024tlc}) \\
\alt{B}{5} & NANOGrav 15\,yr & IPTA DR2 & \cite{Puchades-Ibanez:2026ksh} \\
\bottomrule
\end{tabular}
\caption{The modelling choices compared in figure~\ref{fig:budget}, grouped by the steps of
figure~\ref{fig:pipeline} (N: nucleation rate, P: percolation and reheating, G: GW source, B: binaries
and PTA data), and works that use the alternative. In parentheses: works that compare it with the
default. DMX: piecewise-constant dispersion measures. \label{tab:switches}}
\end{table}

\subsection{Bubble nucleation rate}
\label{sec:alt_rate}

The alternatives for the nucleation rate act on the signal only through $\bHR$, since $\Treh$ is fixed through
$\Delta V$ and the degrees of freedom at reheating, which are practically unchanged when only the bubble nucleation rate
$\Gamma(T)$ is varied.
The phase transition percolates once the exponent $S_{\rm tot}$ has fallen to a threshold around $110$--$160$
for $g = 0.42$ to $0.8$ (see figure~\ref{fig:exponent}), and $\bHR$ follows from the slope of
the exponent there. A correction that shifts $S_{\rm tot}$ by a temperature-independent constant hence merely moves
$\Tp$, but leaves $\bHR$ unchanged up to small per-cent-level shifts.
As we will see below, this observation explains the shifts in the preferred regions of
figure~\ref{fig:budget} within the variants of the nucleation rate.

\paragraph{\alt{N}{1}--\alt{N}{2}: Perturbative uncertainties of the NLO rate.\altanchor{N1}\altanchor{N2}}
In variant \lk{N1}, the four-dimensional matching scale $\mu_4 = 4\pi\eu^{-\gamma_{\rm E}}\,T$ and the
three-dimensional renormalisation scale $\mu_3 = gT$ are both halved and doubled, one after the other. These
variations change $\Tp$ by less than five per cent and leave the preferred region in place. Variant
\lk{N2} concerns the effect of the dynamical prefactor $\Adyn$ of eq.~\eqref{eq:rate} on the normalisation
of the nucleation rate. Langer's prefactor, $\Adyn = \sqrt{|\lambda_-|}/2\pi$~\cite{Langer:1969bc, Affleck:1980ac},
is reproduced on the lattice for a fully thermalised metastable phase~\cite{Hirvonen:2025hqn}, whereas real-time
simulations can find rates lower by about a factor of four~\cite{Gould:2026mtl}.
Here we conservatively vary $\Adyn$ by an order of magnitude in both
directions, which shifts the exponent $S_{\rm tot}$ by $\ln 10 \simeq 2.3$ units. As argued above, this moves $\Tp$
by up to $40\%$ in the preferred interval of $g$, but leaves $\bHR$ constant up to shifts of at most $6\%$.
The centre of the posterior distribution of $g$ hence moves by at most $0.004$.
Since $\Treh$ is unaffected, $v$ does not move either.

\paragraph{\alt{N}{3}: LO instead of NLO.\altanchor{N3}}
In variant \lk{N3} we use the LO instead of the NLO nucleation rate. The LO exponent is larger than the NLO one
by $7$ to $10$ units for a given $g$ and $T/v$ (see figure~\ref{fig:exponent}). At LO the bubble
nucleation rate is hence lower by several orders of magnitude, and the phase transition percolates later. Since
$S_{\rm tot}$ drops more slowly with $T$ at lower temperatures, also $\bHR$ drops, meaning that at LO the phase
transitions are slower. The same $\bHR$ is then reached at a slightly larger coupling, and the preferred interval
moves from $g = 0.43$--$0.50$ to $0.44$--$0.51$.\footnote{Estimating the effect of the
next-to-next-to-leading order by shifting the exponent $S_{\rm tot}$ by $\pm x_3|\delta S|$, with $x_3 =
g_3^2/(\pi m_3)$, between $0.29$ and $0.38$ in the preferred regions, the loop
expansion parameter of the EFT (see appendix~\ref{app:rate}) and $\delta S = S_{\rm NLO} - S_{\rm LO}$, the
preferred region moves even less than between LO and NLO.}

\paragraph{\alt{N}{4}: $V_{\rm 4D}$ + Daisy at the thermal scale.\altanchor{N4}}
In variant \lk{N4} we replace the EFT prediction of the nucleation rate by that of the four-dimensional
daisy-resummed one-loop
potential of eq.~\eqref{eq:V4D} in the $\overline{\rm MS}$ scheme, with the couplings $g$ and $\lambda$ run at
one loop from
$\muz$, where $\lambda(\muz) = 0$, to the thermal scale $\mu = \pi T$, keeping the LO prefactor of
eq.~\eqref{eq:Stot}. Ref.~\cite{Christiansen:2025xhv} chose this scale because there the
action of the critical bubble of the high-temperature expansion of this potential agrees with that of two-loop
dimensional reduction. Between $\muz$ and $\pi T$ the running drives the quartic
coupling negative, to $\lambda(\pi T) \simeq -0.02$ at $g = 0.5$, which lowers the thermal mass of the scalar and
with it the barrier between the phases.
The exponent $S_{\rm tot}$ then lies between the LO and the NLO exponents, about $2$ units below the former
and $6$ to $8$ units above the latter
(see left panel of figure~\ref{fig:exponent}). At $g = 0.5$ the
phase transition percolates at $\Tp = 132$\,keV instead of $210$\,keV, and $\bHR$ drops from $16$ to $15$.
As for the LO comparison,
the same $\bHR$ is reached at a slightly larger coupling, and the preferred interval moves 
from $g = 0.43$--$0.50$ to $0.44$--$0.50$, while $v$ stays in place.
With the same running to $\pi T$ but a one-parameter approximation
of the high-temperature potential~\cite{Levi:2022bzt}, which does not resum the daisy diagrams,
ref.~\cite{Puchades-Ibanez:2026ksh} finds the best-fit gauge coupling larger than with its NLO rate, by $9\%$ for
NANOGrav and by $13\%$ for IPTA DR2, where \lk{N4} moves the preferred interval by about $1\%$. The daisy
resummation does not explain this difference: without the daisy term the exponent of \lk{N4} rises by only $2$ to
$3$ units at $g = 0.5$, while a shift of $g$ by $10\%$ would correspond to about $25$ units (\lk{N5}).

\paragraph{\alt{N}{5}: $V_{\rm 4D}$ + Daisy at the vacuum scale.\altanchor{N5}}
In variant \lk{N5} the Coleman--Weinberg part of eq.~\eqref{eq:V4D} is renormalised at the vacuum scale, $\mu
= v$, with couplings
that do not run, as in refs.~\cite{Balan:2025uke, Goncalves:2025uwh, Bringmann:2026xcx}. The quartic coupling then
stays positive, $\lambda = 0.0044$ at $g = 0.5$, and at a fixed temperature the exponent lies about $30$ units above
that at NLO (see left panel of figure~\ref{fig:exponent}). Renormalising at $\mu = v/2$ or $2v$, i.e.~changing
the renormalisation condition for $\lambda$ from eq.~\eqref{eq:dV}, changes the exponent by less than one
unit, because the
change of scale is absorbed into the quartic coupling. Renormalising at the thermal scale, $\mu = \pi T$, instead
(\emph{not} shown in figure~\ref{fig:exponent}) results in $\lambda \simeq -0.023$ at $T = 10$\,keV and lowers
the exponent by $14$ to $16$ units between $T = 10$\,keV and
$370$\,keV, about $60\%$ of the shift in the exponents $S_\mathrm{tot}$ found in variant \lk{N4}.
Most of the difference between the 4D + Daisy predictions for $S_\mathrm{tot}$ shown in
figure~\ref{fig:exponent} is hence due to the large
logarithms $\ln(T/v)$ that the vacuum scale leaves at the barrier.
The remaining difference comes from the running of $\lambda$, which
at the thermal scale makes it more negative than the Coleman--Weinberg prescription of
ref.~\cite{Coleman:1973jx} predicts, as adopted for deriving eq.~\eqref{eq:dV}.

With $\mu = v$ the bubble nucleation rate is many orders of magnitude
lower at a fixed temperature compared to the NLO prediction, and the percolation temperature drops with
respect to the one predicted by the NLO
computation. For instance, at $g = 0.5$ it drops by over an order of magnitude, from $\Tp = 210$\,keV to $15$\,keV.
Accordingly, at the temperature at which $\bHR$ is evaluated, the exponent $S_{\rm tot}$ has a much shallower slope,
meaning that the bubble number density is lower and $R_\star H$ larger by a factor of up to two. In order to
compensate for this shift in the inference, less supercooling is needed: about $10\%$ larger values of the gauge
coupling are favoured, $g = 0.48$--$0.54$, while the vacuum expectation value $v$ is almost unaffected (see
also the $\Treh$ line in figure~\ref{fig:plane}).
The
distance between the posterior means is hence as large as $\DM = 5.9$. As a result of the
shift in $g$, also the threshold coupling below which the universe stays trapped in the false vacuum
moves about $10\%$ higher,
to $g \simeq 0.43$--$0.46$, which is around the best fit obtained with the NLO computation.

\subsection{Percolation, reheating, and bubble separations}
\label{sec:alt_perc}

Next we discuss alternative choices that change how a given bubble nucleation rate is used to compute the reheating
temperature $\Treh$ and the transition speed $\bH$. As discussed above and illustrated in figure~\ref{fig:plane},
a shift of $\Treh$ can generally be
compensated by a shift of the vacuum expectation value $v$ in the opposite direction, whereas the transition speed
is mainly set by the gauge coupling $g$, such that any influence on $\bH$ can be compensated by a shift in $g$.

\paragraph{\alt{P}{1}: Expansion history during the transition. \altanchor{P1}}
Variant \lk{P1} studies the impact of the redshift of the energy density within the bubbles throughout the
initial stages of the transition.
It approximates the Hubble rate as $3M_{\rm Pl}^2H^2 = (1 - P_{\rm t})\,\rho_{\rm f} + P_{\rm
t}\,\rho_{\rm t} \simeq \rho_{\rm f}$ (see eq.~\eqref{eq:friedmann}), i.e.\ through the assumption that the energy
density in the false vacuum, dominantly sourced by $\Delta V$, is approximately equal to that in the broken phase.
Until percolation most of the volume is still in the false vacuum, and the converted regions hold the released energy
as radiation that has redshifted only little, so the two expansion rates hardly differ.
In turn, we find that the effect on the
preferred gauge coupling and vacuum expectation value is negligible.

\paragraph{\alt{P}{2}: Percolation threshold.\altanchor{P2}}
Another negligible impact on the preferred parameter space comes from the choice of the percolation threshold
$\fperc$ at which the percolation temperature $\Tp$ is defined.
The default value $\fperc = 0.29$ is based on Monte Carlo simulations under idealised
conditions~\cite{Rintoul:1997tze, Lorenz:2001}.
In other studies, a different value of $\fperc$ is used.
Here we compare with ref.~\cite{Lewicki:2025hxg} and set $\fperc =
1 - 1/\eu \simeq 0.63$~\cite{Enqvist:1991xw}.
The threshold enters both the bubble number density $\nb$, through the
temperature $\Tp$ at which it is evaluated, and the transition speed $\bHR$ of eq.~\eqref{eq:speeds}, through factors
$\fperc^{1/3}$ and $\fperc^{-1/3}$ that cancel for an exponential nucleation rate.
Also in the present case the effect on
the preferred parameter space is found to be negligible.

\paragraph{\alt{P}{3}: Require decreasing $a^3(1 - P_{\rm t})$ at $\Tp$.\altanchor{P3}}
In ref.~\cite{Ellis:2018mja}, based on earlier work~\cite{Turner:1992tz}, a proxy for the completion of the phase
transition has been argued for, which requires that the physical volume of the false vacuum, $\mathcal{V}_{\rm
f} \propto
a^3(1 - P_{\rm t})$, already decreases at percolation, i.e.\ that
\begin{equation}
  \frac{1}{H}\,\frac{\dd\ln\big(a^3(1 - P_{\rm t})\big)}{\dd t}\bigg|_{t_{\rm p}} = 3 - \frac{1}{H}\,\frac{\dd I}{\dd t}\bigg|_{t_{\rm p}} < 0\,.
  \label{eq:local}
\end{equation}
We find that for the slowest phase transitions of the NANOGrav-preferred region of parameter space, to the
left of the orange line in figure~\ref{fig:plane}, the physical false-vacuum volume still grows at percolation. There,
however, the phase transition can still complete nonetheless, because the bubble nucleation rate keeps rising after
percolation, such that the conversion soon outpaces the expansion of space between the bubbles and the true-vacuum
fraction reaches the threshold $0.99$ within $0.3$ to $0.9$ e-folds after percolation. Applying this criterion 
would thus discard phase transitions that complete and remove the slowest ones, below $g \simeq 0.45$, and cut the
preferred region to $g = 0.46$--$0.51$ and $v \in [0.81, 2.3]$\,GeV.

\paragraph{\alt{P}{4}: $\bHS$ instead of $\bHR$.\altanchor{P4}}
In variant \lk{P4} we replace both $\bHR$ and $\bHG$ in the GW templates by the slope $\bHS$ of the exponent
$S_{\rm tot}$, see eq.~\eqref{eq:speeds}.
Since $\bHS$ is larger than $\bHR$ ($\bHG$) at every parameter point, typically by about $20\%$ ($30\%$), and
the phase transitions
become slower for smaller $g$, the interval moves down to $g = 0.43$--$0.48$.
The preferred region in this variant is not cut off at low $g$ as could be expected based on figure~3 of
ref.~\cite{Matuszak:2026xsz} which shows that $\bHS<0$ is possible and generic in the region of strongest
supercooling realisable in a model with a polynomial potential. The underlying reason for this somewhat
surprising result is also illustrated in figure~9 of ref.~\cite{Bringmann:2026xcx}: as opposed to other
models, the present one always features at least a very shallow barrier even at very low temperatures, such that in
the present model $\bHS$ is always positive.

\paragraph{\alt{P}{5}: Instantaneous reheating.\altanchor{P5}}

The reheating temperature $\Treh = T_{\rm b}(\Tp)$, i.e.~the broken-phase temperature evolved to percolation
according to eq.~\eqref{eq:Tb}, is often approximated in the instantaneous-reheating approximation (see for
instance refs.~\cite{Basler:2024aaf,Costa:2025pew,Madge:2023dxc}),
\begin{equation}
  T_{\rm reh} \simeq \left(\frac{g_{\rm eff}(\Tp)}{g_{\rm eff}(T_{\rm reh})}\right)^{1/4} (1+\alpha)^{1/4}\, \Tp \approx (1+\alpha)^{1/4}\, \Tp\,,
  \label{eq:Trehapprox}
\end{equation}
which follows from energy conservation in that limit.
In the second step, the number of degrees of freedom at reheating in the broken phase is approximated by that
in the symmetric phase at percolation.
This approximation overestimates $\Treh$ for two reasons:
\emph{i)} there are more relativistic degrees of freedom in the plasma after reheating, typically $g_{\rm
eff}(\Tp) = \mathcal{O}(5$--$10)$ and $g_{\rm eff}(\Treh) = \mathcal{O}(20$--$50)$ including those of the dark
sector, and \emph{ii)} the radiation in the transitioned regions already redshifts before percolation, which
lowers $\Treh$ by up to $20\%$. Together, both effects lead to an overestimate of $\Treh$ by a factor of $1.2$ to
$2.1$ when following eq.~\eqref{eq:Trehapprox}, which can be compensated by a downward shift of $v$ by the same factor.
In this variant the region hence moves
down to $v \in [0.69, 2.7]$\,GeV, with $g$ almost unaffected.

\subsection{Gravitational wave source}
\label{sec:alt_source}

Even for fixed reheating temperatures $\Treh$ and transition speeds $\bHR$, the uncertainty related to mapping
them to specific GW signatures is large, as indicated by the spread of GW templates in figure~\ref{fig:spectra}.
The respective shifts between these templates, all evaluated at the best fit in $g$ and $v$ of the default
computation, allow for an intuition on their effect within the inference problem: Generally speaking, we find
that a larger-than-observed spectral amplitude can be compensated by a faster phase transition, i.e.\ a
larger $g$, and a higher peak frequency by a smaller $v$. A steeper rise of the spectral template at frequencies
below the peak can also be compensated by a smaller $v$ in the fit, at least up to some degree, since the
spectral turnover just below the peak
then shifts to lower frequencies, flattening the spectrum just below it.
The latter is allowed by the data since the high-frequency bins carry little statistical weight, as indicated
by their corresponding free-spectral densities having wide, almost uninformative posteriors, rather
representing an upper bound on the peak amplitude, see figure~\ref{fig:spectra}.

We want to emphasise again that \textit{none} of the templates discussed below was fitted to a simulated
phase transition with a bubble nucleation rate like the one at hand.
Instead a purely exponential nucleation rate is assumed, which is a good approximation only in the limit of
large $\beta/H$ (see section~\ref{sec:source}).
Moreover, none of the spectra comes with a systematic uncertainty estimate.
The following discussion is thus limited to the discussion of the impact of the \emph{different templates} on
the preferred parameter space, but does not allow conclusions to be drawn on the uncertainties associated with
the \emph{templates themselves}.
The latter would require a dedicated study, which is beyond the scope of this work.

\paragraph{\alt{G}{1} and \alt{G}{2}: No expansion between bubbles, free-streaming shells.\altanchor{G1}\altanchor{G2}}
Within the template of ref.~\cite{Lewicki:2025hxg} which we treat as a default here, the expansion of space
during the transition is taken into account and the \emph{dissipative} bulk-flow model is assumed, which is
valid for the case of shells decaying quickly after their collisions.
Variant \lk{G1} switches off the expansion and instead uses the fit of ref.~\cite{Lewicki:2022pdb} in order to
separate the effect of
expansion on the inference.
Within this template, $A = 0.051$ and $2\pi f_{\rm p}/\beta = 0.77$, independent of $\bHG$ (compare eq.~\eqref{eq:lv}).
These simulations have the same scalar-field shells whose
energy-momentum tensor falls with the radius as $R^{-3}$ as the dominant GW source.
As opposed to their description in ref.~\cite{Lewicki:2022pdb}, we keep the causal $f^3$ tail of the spectrum
below $f_{\rm H} = a_{\rm p}H_{\rm p}/(2\pi a_0)$ of eq.~\eqref{eq:redshift}, in order to separate the effect
of expansion alone and not confound it with the effect of an alternative deep-infrared behaviour of the
spectrum.\footnote{\label{fn:causal}We checked the
effect of omitting the causal factor within all fits of ref.~\cite{Lewicki:2022pdb} without expansion.
Without it, i.e.~taking the fits at face value given there, they rise below $f_{\rm H}$ with their own slope
$n_{\rm l} = 1.0$ to $3.1$.
To compensate for the shallow slope, the upper ends of the $1\sigma$ intervals of $v$ then move up to $9$ to
$17$\,GeV for the fits with an $R^{-3}$ decay of the energy-momentum tensor, and their $2\sigma$ intervals up
to $32$ to $63$\,GeV.
The fits with an $R^{-2}$ decay require even larger vacuum expectation values and reach the upper end of the
prior range at $v = 100$\,GeV, featuring a maximum $\ln L$ lower by about $2$ units, signalling a notably worse
fit to the data.} The resulting spectrum is shown in figure~\ref{fig:spectra} as a blue line.
There it appears very close to the default spectrum (thick black line), since at the best fit $\bHG = 7.7$
and $\bHR = 9.6$, such that the spectral parameters $A$, $f_\mathrm{p}$ and $f_\mathrm{H}$ are close to their
default prediction of eq.~\eqref{eq:lv}.
The spectral slopes of this template, however, lie at slightly different values of $n_\mathrm{l} = 2.41$,
$n_\mathrm{h} = 2.42$, and $w = 4.08$ (see table~1 in ref.~\cite{Lewicki:2022pdb}) as opposed to the
$n_\mathrm{l} = n_\mathrm{h} = w = 2$ in our default template, accounting for the visible differences.
For slower transitions, the differences between the templates become more pronounced.
Due to the missing suppression factor within $A$, the thin spike-shaped region at small $g$ visible in
figure~\ref{fig:plane} then vanishes,
and in turn the lower end of the $2\sigma$ interval of $g$ is shifted up by roughly 6\%.
Moreover, the preferred region shrinks towards slightly smaller $v$.
Keeping the expansion of space but
instead assuming shells that keep their energy and propagate freely after the collisions (variant \lk{G2},
orange curve in figure~\ref{fig:spectra}) follows
eqs.~\eqref{eq:bpl} and~\eqref{eq:lv} with $(n_{\rm l}, n_{\rm h}, w) = (1, 2, 1)$~\cite{Lewicki:2025hxg}.
Compared to the default, this spectrum is shallower, and thus the region moves up to $v \in [1.8, 8.0]$\,GeV
in order to compensate for the shallower slope.
In this case the posterior of
$g$ remains close to identical to the one with the default template.
This is due to the spectral amplitude being identical to the default template. 

\paragraph{\alt{G}{3} and \alt{G}{4}: Sound-wave templates.\altanchor{G3}\altanchor{G4}}
A family of spectral templates that are often used as defaults are those recommended by the LISA cosmology
working group in ref.~\cite{Caprini:2019egz} from 2019 and in ref.~\cite{Caprini:2024hue}
from 2024. The 2019 template relies on hybrid simulations of the bulk fluid motion
and the scalar field of the bubble walls~\cite{Hindmarsh:2017gnf}, which are limited in resolution and in their reach
in $\alpha$.
The improved 2024 template instead relies on higgsless simulations~\cite{Jinno:2022mie}, which do not evolve the
bubble walls dynamically and can hence simulate the fluid motion more efficiently. Nevertheless, also these templates
describe the physics faithfully only up to moderate supercooling, $\alpha \lesssim 1$, for walls that do not run away
and a plasma that absorbs the released energy. The case at hand is in the opposite limit, $\alpha \gg 1$ with highly
relativistic bubble walls, so none of these assumptions holds here (see also discussion in section~\ref{sec:source}).
In the variants \lk{G3} (higgsless, 2024) and \lk{G4} (hybrid, 2019) we compare the impact of these templates
on the preferred parameter space.
Both are written in terms of the
bubble separation, which we evaluate as $(RH)_{\rm LISA} = (8\pi)^{1/3}/\bHR$, and of the kinetic energy fraction of the
fluid, $K = \kappa_{\rm sw}\alpha/(1+\alpha)$, with the efficiency $\kappa_{\rm sw} = \alpha/(0.73 +
0.083\sqrt{\alpha} + \alpha)$ for luminal walls~\cite{Espinosa:2010hh}.
For the 2024 template, a multi-bubble correction of $\kappa_{\rm sw} \to 0.6 \kappa_{\rm sw}$ to the
single-bubble fit of the kinetic-energy efficiency is multiplied~\cite{Jinno:2022mie, Caprini:2024gyk}.
The source lasts $\min(1,
H\tau_{\rm sh})$ Hubble times, with $\tau_{\rm sh}$ the time in which the fluid develops shocks~\cite{Guo:2020grp},
the sound speed is $\cs = 1/\sqrt3$, that of a radiation fluid, and no suppression due to Hubble expansion as
in eq.~\eqref{eq:lv} or a continuation to an $f^3$ causal slope is applied.
A notable difference to the default spectra moreover lies in the spectra only scaling with
$\left(\bH\right)^{-1}$ instead of $\left(\bH\right)^{-2}$ as in eq.~\eqref{eq:bpl}.
For a given $\bHR$, the spectra are hence peaking at a larger amplitude compared to the default spectra from
ref.~\cite{Lewicki:2025hxg}, about an order of magnitude higher for the case illustrated in
figure~\ref{fig:spectra} with $\bHR = 9.6$ (curves in magenta and purple).
The shift towards frequencies higher by a factor of a few is due to the relevant length scale of the
transition modelled through the thickness of the sound shells found in the simulations, which is smaller than
the mean bubble separation~\cite{Hindmarsh:2017gnf, Caprini:2024hue}. Both effects push the region the same way, towards
faster phase transitions at smaller $v$: the higgsless template prefers $g = 0.50$--$0.60$ and $v \in [0.10,
0.77]$\,GeV, with an unchanged goodness-of-fit, and the hybrid template $g = 0.54$--$0.62$ and $v \in [0.04,
0.16]$\,GeV with a slightly worse fit ($\Delta \ln L = - 0.4$).
Due to the lower favoured vacuum expectation values the reheating temperature approaches its lower bound,
$\Treh > 6$\,MeV~\cite{Barbieri:2025moq}, which excludes $4\%$ and $42\%$ of the posterior volume for the 2024
and the 2019 template, respectively.
We repeat that both templates are \emph{not} applicable in the preferred regions, since they assume a plasma
that absorbs the released energy, which is not the case here.
The results of this comparison are hence only indicative of the effect of the different templates on the
preferred parameter space and of the effect of using more refined templates even for the same signal source.

\paragraph{\alt{G}{5}: Envelope approximation.\altanchor{G5}}
Variant \lk{G5} studies the envelope approximation, a template often used throughout the literature as it was
one of the first to be derived from simulations of colliding bubbles~\cite{Kosowsky:1992vn, Huber:2008hg,
Jinno:2016vai}.
It is a predecessor of the bulk-flow model on which the default template is based.
It assumes that the energy-momentum tensor of the bubble walls vanishes after they collide, and keeps only the
envelope of the uncollided walls.
In the description of ref.~\cite{Lewicki:2025hxg} it has a spectrum rising as $f^{3.1}$ and a peak frequency
lying a factor of about two higher than that of the default bulk-flow spectra, see the brown curve in
figure~\ref{fig:spectra}.
The envelope spectrum rises slightly steeper than the default template, thus moving the preferred region to
smaller $v \in [0.31, 1.7]$\,GeV corresponding to a posterior distance $\DM = 4.7$.
The goodness-of-fit is slightly worse than for the default template, with a decrease of $\ln L$ by $0.4$ units.
As discussed in footnote~\ref{fn:causal}, whether the causal factor is included in this template or not does
not impact the parameter inference since its $n_{\rm l} = 3.1$ slope is close to the $f^3$ behaviour of the
causal tail.

\subsection{Pulsar timing data, noise models, and binary background}
\label{sec:alt_data}

The previous alternatives left out a dominant source of uncertainty in the inference, i.e.~the binary background.
A binary background adds a contribution to the spectrum, typically with a shallow slope
and with an amplitude several orders of magnitude below the one inferred
from the PTA data.
Both its slope and amplitude, however, are under active debate~\cite{Sato-Polito:2023gym, Sato-Polito:2025ivz,
Chen:2025wel}.
Depending on the specific binary hardening model, galaxy merger rates, and black hole masses, it can also take
over part of the signal or even explain it fully.
Moreover, our use of the free spectral likelihood, which already assumes a specific noise model for the
pulsars themselves, as well as the modelling of the propagation of radio signals throughout the galaxy, can
also affect the inference.
Also the choice of the PTA data influences the analysis.
In this section we focus on the NANOGrav 15\,yr data set, which offers the largest variety of publicly
available likelihood products and hence allows us to study the impact of the different modelling assumptions
on the inference stated above, and eventually compare to the IPTA DR2 data set.

\paragraph{\alt{B}{1}: Treatment of the binary background.\altanchor{B1}}
A simple
estimate of the binary signal, which is typically used in PTA new-physics searches~\cite{EPTA:2023xxk,
Madge:2023dxc, Figueroa:2023zhu, Ellis:2023oxs, Wang:2025wht, Bringmann:2023opz, Balan:2025uke, Bringmann:2026xcx,
Domcke:2026xza} is a power law $h^2\OGW \propto A_{\rm bhb}^2 (f/f_{\rm yr})^{5-\gamma_{\rm bhb}}$ with an
amplitude $A_{\rm bhb}$ at the pivot frequency $f_{\rm yr} = 1/{\rm yr}$ and a spectral index $\gamma_{\rm bhb}$.
Also the new-physics searches of NANOGrav use this parametrisation and recommend a prior set on $(A_{\rm bhb},
\gamma_{\rm bhb})$~\cite{NANOGrav:2023hvm}.
To arrive at these priors, an instantaneous binary hardening is assumed.
Elsewhere~\cite{NANOGrav:2023hfp}, NANOGrav argue that this parametrisation \emph{is not self-consistent, as
GW-only evolution is unable to bring binaries to the PTA band within a Hubble time}.
A better description of the binary signal is the population model of \texttt{binaryfog} introduced in
section~\ref{sec:data}.
In variant \lk{B1} we compare the two descriptions: the population under each of its three prior sets of
table~\ref{tab:binpriors}, and the power law under the priors recommended in ref.~\cite{Mitridate:2023oar}.
In either
case we marginalise the likelihood over the astrophysical parameters to arrive at the shown marginalised
posteriors of $g$ and $v$.

We find that, independently of the prior set and functional form of the binary contribution, the posteriors of
$g$ and $v$ widen, mostly towards larger $g$ (i.e.~weaker cosmological contributions).
In turn, also the 
posterior mean shifts. Whereas the power law takes over
part of the signal amplitude at high frequencies and widens the posterior intervals only slightly, the effect of the
population-based backgrounds is generally larger.
By how much the parameter space opens up depends on the imposed prior set.
Under the survey-averaged priors, recommended in ref.~\cite{Bringmann:2026uxo}, the $2\sigma$ interval reaches
$g = 0.7$, and the posterior mean shifts from $g = 0.47$ to $0.49$, which raises $\bHR$ by about $15\%$.
The other two prior sets, both based on the same galaxy survey data and one version of the black hole--bulge
mass relation (but including a somewhat artificially widened prior width in the case of the NANOGrav
prior~\cite{Bringmann:2026uxo}), predict a smaller contribution of the binaries to the measured amplitude.
Hence, the posterior widens less in these cases.
The posterior obtained under the survey-averaged prior set is shown in figure~\ref{fig:plane} in dashed contour lines.

We point out that the
ranges illustrated in figure~\ref{fig:budget} are Bayesian credible intervals.
With the binaries marginalised, they hence depend on the adopted prior ranges of $g$ and
$v$: where the phase transition contributes little, the binaries alone explain the signal, and the posterior keeps a
plateau of $1\%$ to $7\%$ of its maximum up to the edges of the prior range, whose weight grows with the range.
The posterior volume hence scales with the prior volume.
Variant \lk{B1} uses the prior ranges of every other variant, $0.39 < g < 0.75$ and $10\,{\rm MeV} < v <
100\,{\rm GeV}$ (section~\ref{sec:preferred}), which contains the NANOGrav posterior of every variant of
figure~\ref{fig:budget} without binaries.
Widening the range of $v$ to $10\,{\rm MeV} < v < 10\,{\rm TeV}$, the distance of the posterior means grows
from $\DM = 2.9$ to $5.5$ under the survey-averaged priors (from $1.9$ to $3.9$ under the NANOGrav priors),
the $2\sigma$ interval of $v$ extends up to about $1$\,TeV, and that of $g$ up to $0.71$ instead of $0.69$.
In appendix~\ref{app:profiling} we show the results of the corresponding frequentist analysis of the profiled
likelihood, circumventing the issue of prior dependence, though requiring a different interpretation of the
obtained contours.

\paragraph{\alt{B}{2}: Hellings--Downs instead of CURN.\altanchor{B2}}
The free spectral likelihood of the NANOGrav 15\,yr data set exists in several versions which assume different
correlations of the signal between pulsars. The signal was first identified as common uncorrelated red noise
(CURN)~\cite{NANOGrav:2020bcs}. In the 15\,yr data set, the evidence for this red noise to be Hellings--Downs
correlated, i.e.\ due to a GW background, increased to between $3\sigma$ and $4\sigma$, depending on the used
test statistic~\cite{NANOGrav:2023gor}.
The Hellings--Downs
version of the free-spectrum likelihood attributes all of the common red noise that remains after subtracting
the pulsar-intrinsic red noise from the pulsar timing residuals to a GW background.
Variant \lk{B2} uses this free spectrum instead of CURN.
We find that the posterior region hardly
moves, though it widens slightly towards larger $g$, i.e.~faster transitions and marginally weaker GW signals,
which is due to the HD free spectrum allowing a wider range of GW amplitudes due to the more limited
statistical power of it which motivated us to use the CURN version as a default of the present analysis in the
first place.

\paragraph{\alt{B}{3}: DMGP instead of DMX.\altanchor{B3}}
The chromatic delays from the dispersion of the radio pulses in the interstellar medium are fitted as deterministic,
piecewise-constant dispersion measures (DMX) by default in the analyses of NANOGrav. Alternatively, NANOGrav also
presents its results with the chromatic delays modelled stochastically as a Gaussian process (DMGP), which is argued
to describe the delays better, given the stochastic nature of the variations in the interstellar
medium~\cite{NANOGrav:2023gor, NANOGrav:2023ctt}. The analysis with this noise model was not repeated for the 2026
re-release of the free-spectrum likelihoods, so in variant \lk{B3} we use the 2023
release~\cite{NANOGrav:2023kde, PTArcadeData}.\footnote{We thank William Lamb for a private version
of the DMGP likelihood updated with the 2026 fixes, which shows no relevant changes to the results derived
with the 2023 version used here.} With DMGP the region moves by about its width to faster
phase transitions at smaller $v$, $g = 0.48$--$0.58$ and $v \in [0.18, 1.6]$\,GeV, the direction expected if the
recovered spectrum rises more steeply. This direction agrees with ref.~\cite{Agarwal:2026dlv}, whose chromatic noise
models customised for each pulsar recover a steeper spectrum and move the phase-transition fit of NANOGrav to lower
temperatures. No free spectrum based on such models~\cite{Agarwal:2026dlv, Larsen:2026pru} has been released yet.

\paragraph{\alt{B}{4}: Number of frequency bins.\altanchor{B4}}
By default, we use $14$ bins of the NANOGrav free-spectral likelihood, the choice of the new-physics search of
NANOGrav~\cite{NANOGrav:2023hvm}.
In variant \lk{B4} we replace these with $5$ and $30$ frequency bins.
One reason why higher frequency bins
are usually removed from the analysis is that they are dominated by white noise, i.e.\ that they do not contain
relevant spectral information on the GW background. Above the fourteenth bin of the NANOGrav 15\,yr data set the
white noise of the pulsars dominates and the data are consistent with no signal, so $30$ bins leave the region in
place.
With $5$ bins, the default range of the outdated NANOGrav 12.5\,yr analysis~\cite{NANOGrav:2020bcs}, the fit loses the
frequencies at which the spectrum turns over, the peak frequency is less constrained, and the region widens towards
faster phase transitions at smaller $v$, $g = 0.46$--$0.55$ and $v \in [0.26, 2.6]$\,GeV.

\paragraph{\alt{B}{5}: IPTA data set.\altanchor{B5}}
Finally, we compare the NANOGrav-preferred regions to those obtained with the IPTA DR2 data
set~\cite{Antoniadis:2022pcn}.
Here we only replace the NANOGrav data set with the IPTA data set and leave the full comparison of each of the
above variants with the IPTA data set for the appendix, as the results are qualitatively similar, see
table~\ref{tab:switch_numbers_ipta} in the appendix.
The most relevant shift is the preference for smaller vacuum expectation values $v$ compared to the NANOGrav data set.
As can be anticipated from figures~\ref{fig:plane} and \ref{fig:spectra}, the posterior in $v$ lies a factor
of $3$ to $6$ below the NANOGrav preferred values, and the gauge couplings are also larger by about
$2$--$11\%$, in every variant (except those with $5$ or $30$ frequency bins).
Both observations are consistent with the IPTA free spectrum turning over at a lower frequency, around
$10$\,nHz, and having a smaller amplitude than the one inferred by NANOGrav (see also
section~\ref{sec:preferred}).

\subsection{Does NANOGrav observe a dark sector phase transition?}
\label{sec:alt_binaries}

Before summarising the results of the previous subsections, we want to discuss the significance with which the
NANOGrav 15\,yr data favour a contribution of a phase transition over the binaries alone.
Since the preference depends on the goodness-of-fit of the combined fit, which can only increase compared to
the fit of the binaries alone, and since the latter is not impacted by the modelling choices discussed above
(see table~\ref{tab:switch_numbers} in the appendix), here we stick with the default modelling choices
presented in section~\ref{sec:signal}.
How much of the signal spectrum can be attributed to the binaries, however, depends on the adopted
astrophysical priors, and so does the significance with which the data favour a contribution of the phase
transition over the binaries alone.
As described in appendix~\ref{app:inference} in more detail, we measure the local statistical preference by
comparing the maximum of the sum of likelihood and astrophysical prior $\ln L + \ln \pi$ over $g$, $v$, and
the astrophysical parameters, minus the same
maximum when considering only the binary contribution alone.

With the survey-averaged priors, we find a very mild preference for a phase transition with a significance of
$2.1\sigma$.
Under the NANOGrav priors, having wider prior widths, the preference drops slightly to a hint of $1.9\sigma$,
and with the Tomczak--Kormendy priors, which predict a smaller binary contribution, it rises to a notable
$3.6\sigma$ preference.
We note that these values are slightly higher than the ones given in ref.~\cite{Bringmann:2026uxo}, which were
obtained for a phase transition described by three input parameters ($\alpha$, $\bHR$, and $\Treh$) instead of
the two parameters $g$ and $v$ used here, based on the Hellings--Downs free spectra instead of the CURN ones,
and with the higgsless sound-wave template of variant \lk{G3} instead of the dissipative bulk-flow template used as
a default here~\cite{Lewicki:2025hxg}.

\begin{figure}[t]
\centering
\includegraphics[width=0.6\textwidth]{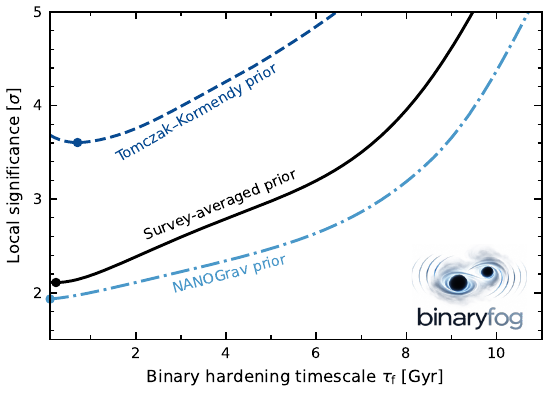}
\caption{Local significance within the NANOGrav 15\,yr data for a phase transition contribution on top of a
binary signal as modelled through
\texttt{binaryfog}~\cite{Bringmann:2026uxo}.
For increased binary hardening timescales $\tau_{\rm f}$ of the binaries, the preference for a contribution increases
since the binary signal weakens.
The three lines correspond to the three astrophysical prior choices listed in table~\ref{tab:binpriors}.
Dots indicate the preference for a phase transition at the best fit of the binaries alone, i.e.~the minimum of
each shown curve.
\label{fig:hardening}}
\end{figure}

The significances stated above are obtained under the assumed flat priors of $\tau_{\rm f}$ and $\nu_{\rm
inner}$, which are the phenomenological parameters that describe the hardening of the binaries.
As shown in ref.~\cite{Bringmann:2026uxo} the influence of these on both the amplitude and the slope of the
predicted binary contribution is large and hence also the significance for a phase transition contribution
scales with their prior ranges.
The observational evidence for any specific value of these parameters is very limited: In spherical galaxies
the binary ejects the stars on orbits that reach
it faster than two-body relaxation replaces them, and it stalls at parsec separations, dubbed the final-parsec
problem~\cite{Begelman:1980vb, Milosavljevic:2002bn, Merritt:2004gc}. In simulations of merging and of triaxial
galaxies, stellar scattering keeps the binary shrinking and drives it to coalescence within about a
Gyr~\cite{Khan:2011gi, Vasiliev:2015uzo, Sesana:2015haa}, and torques from gas~\cite{Armitage:2002uu} or dynamical
friction in spikes of self-interacting dark matter~\cite{Alonso-Alvarez:2024gdz} could shorten this further. None of
these hardening mechanisms is established for the massive binaries that dominate the PTA band, so there is
good reason to consider different prior ranges as well.

Figure~\ref{fig:hardening} shows the preference at fixed
$\tau_{\rm f}$, where the other five astrophysical parameters, and $g$ and $v$ in the fit with the phase
transition are varied to maximise
$\ln L + \ln \pi$.
The binaries alone fit the NANOGrav data best when short hardening timescales are possible: the best-fits lie
towards the lower
edge of the prior, $0.2$\,Gyr, with the survey-averaged priors, and at $0.1$\,Gyr and $0.7$\,Gyr with the NANOGrav and
the Tomczak--Kormendy priors, where the preferences take the values $2.1\sigma$, $1.9\sigma$, and $3.6\sigma$
quoted above, respectively.
When requiring longer hardening timescales only binaries from the oldest galaxy mergers reach the PTA band
within the age of the universe.
The amplitude of the binary background hence drops, the binaries alone fit the data worse, and the preference for a
contribution of the phase transition rises.
With a lower prior boundary $\tau_{\rm min}$ of the hardening timescale, only the part of the figure to the
right of $\tau_{\rm min}$ remains.
The preference is then the value of each curve at $\tau_{\rm min}$: $2.8\sigma$, $2.3\sigma$, and
$4.3\sigma$ for $\tau_{\rm min} = 4$\,Gyr, and $3.5\sigma$, $2.9\sigma$, and $5.2\sigma$ for
$\tau_{\rm min} = 7$\,Gyr, for the survey-averaged, the NANOGrav, and the Tomczak--Kormendy priors, respectively.
The range of
$1.9\sigma$ to $3.6\sigma$ stated above thus presumes that the binaries can harden within about a Gyr, and the
preference rises if they harden more slowly.

The question asked in the title of this subsection hence cannot be answered without a better understanding of
the astrophysical processes that drive the binaries to coalescence.
Given the current uncertainties, the data are consistent with a contribution of a phase transition, and there
is even a slight preference for it, but the available data do not require it.
With increased data quality, more precise PTA observations, and better astrophysical modelling of the binary
hardening, it will be possible to answer the question more definitively.

\subsection{Summary}
\label{sec:summary}

Taken together, the alternatives show which steps of the computation limit the inference of the parameters of the
dark sector from the PTA data: The mapping of the phase transition to a GW spectrum limits it most. The choice of the
spectral template leaves the vacuum expectation value $v$, which sets the peak frequency, uncertain by orders of
magnitude, and the gauge coupling $g$ moves between the templates for highly boosted fluid shells and the spectra
sourced by sound waves. The noise model of the pulsars, the share of the binaries, and the reheating prescription each
move the preferred region by two to three standard deviations. The binaries in addition decide the significance of a
phase-transition contribution, through their astrophysical priors and their hardening timescale. The nucleation rate,
once computed at NLO in the three-dimensional EFT, limits the inference least. With the
couplings fixed at the vacuum scale within the four-dimensional daisy-resummed potential, as in earlier
fits~\cite{Balan:2025uke,Bringmann:2026xcx}, the same rate moves it by several standard deviations, more than
the noise
model, the binaries, or the reheating prescription.

The small uncertainty of the nucleation rate is a success of the three-dimensional EFT. The renormalisation-scale
dependence and the choice of renormalisation scheme of the four-dimensional rate dominate the reconstruction of the
parameters of classically conformal models from a future LISA signal~\cite{Lewicki:2024xan, Kierkla:2026bnm}, and the
choices of the transition temperature, of the speed of the phase transition, and of the energy budget change the
predicted spectra of electroweak phase transitions by up to three orders of magnitude~\cite{Guo:2021qcq,
Athron:2023rfq}. With the NLO rate of the EFT the uncertainty of the nucleation rate falls below that of every later
step, at least in this simplest classically conformal model, so that these later steps now decide the precision of the
inference, giving a clear target for future work.

\section{Conclusions}
\label{sec:conclusions}

We have followed the PTA signal of a classically conformal dark $U(1)$ from the bubble nucleation rate through
percolation, reheating, and the GW spectrum to the NANOGrav and IPTA likelihoods, and repeated the fit for $20$
alternative modelling choices in use in the literature. With the most complete treatment at each step, the NANOGrav
data prefer gauge couplings $g$ of about $0.43$ to $0.50$, dark photon masses around a GeV, and reheating
temperatures of about $0.1$\,GeV (see figure~\ref{fig:plane}).

The uncertainty of this inference is dominated by the mapping of the phase transition to a GW spectrum. No template
was fitted to simulations of a phase transition as slow and as strong as this one, and the templates for highly
boosted fluid shells leave $v$ uncertain at $1\sigma$ by more than an order of magnitude, and by more than two
orders of magnitude
once spectra sourced by sound waves are included.
The noise model of the pulsars, the binary background, and the reheating prescription
each shift the preferred region by two to three posterior standard deviations. At NLO in the three-dimensional EFT the
nucleation rate is no longer a limiting step, and the rate of the four-dimensional daisy-resummed potential with
running couplings leads to nearly the same inference.
With the couplings fixed at the vacuum scale, as in earlier fits, this
rate shifts the inferred gauge coupling $g$ by about $10\%$, i.e.~several posterior standard deviations.
Whether the data prefer a contribution of the phase transition over the
binaries alone depends on how galaxy survey data is combined to an astrophysical prior and on the details of
binary hardening, with local significances
between $1.9\sigma$ and $3.6\sigma$ for different prior choices.

Several lessons follow for the prediction of GW signals from phase transitions and for their inference from data.
The method used at each step should be stated precisely, from the computation of the nucleation rate and the
definition of the speed of the phase transition, to the template, the noise model of the pulsars and the
astrophysical priors of the binaries, since choices that look like details move the inferred parameters by
several standard deviations.
Public codes for single steps and for the whole chain~\cite{Ekstedt:2022bff, Ekstedt:2023sqc,
Wainwright:2011kj, Athron:2020sbe, Ekstedt:2024fyq, Mitridate:2023oar, Bringmann:2026uxo, Matuszak:2026xsz}
make such results comparable.
The largest uncertainty, the mapping of a temperature-dependent nucleation history $\Gamma(T)$ to a GW
spectrum, needs templates with quantified error bands, which for strongly supercooled phase transitions
requires simulations of highly boosted shells, with the expansion of the universe included.
PTA searches for new physics should be cross-checked with the DMGP noise model and with different priors for
the binaries.
Finally, only a computation fast enough to be repeated for every relevant alternative can quantify the
uncertainty within the inference problem.
We expect the same conclusions derived here to apply to other phase transitions and to upcoming new-physics
signal searches at LISA, clearly indicating that follow-up work is needed to use the full discovery potential
of present and future GW data.

For the present model the next step is a global fit in which the astrophysical parameters of the binaries and the
parameters of the dark sector are inferred at the same time, the modelling choices are varied together, and the
constraints complementary to the PTA data are included: on the kinetic mixing that lets the dark photon
decay~\cite{Rescigno:2025ong, Caputo:2025avc, Caputo:2026pdw}, on a decay channel of the dark
Higgs~\cite{Fradette:2017sdd, Bringmann:2026xcx}, on the reheating temperature~\cite{Hasegawa:2019jsa,
Barbieri:2025moq}, on isocurvature~\cite{Elahi:2026ldn} and curvature perturbations~\cite{Lewicki:2024sfw,
Franciolini:2025ztf, Bringmann:2025cht}, and on the portal that thermalises the two sectors~\cite{Balan:2025uke}.
The publicly available codes \TL\ and \texttt{binaryfog} are ready to use for these next endeavours.

\acknowledgments

I thank Cristina Puchades-Ib\'a\~nez and Pedro Schwaller for sharing the tables underlying
ref.~\cite{Puchades-Ibanez:2026ksh} and for the discussions that made the comparison of
appendix~\ref{app:rate} possible, William Lamb for a private version of the updated
DMGP likelihood, Jonas Matuszak for the joint development of \TL\ and its
percolation solvers, Maciej Kierkla for explaining the hybrid regime to me, Peter Athron for discussions on
the employed GW templates and the transition completion criteria, as well as Chris Gordon and Torsten Bringmann
for their involvement in the \texttt{binaryfog} development and discussions on statistics.
Moreover, I thank Jorinde van de Vis, Jordy de Vries, Marieke Postma~\cite{vandeVis:2025efm} and William Shakespeare~\cite{Macbeth} for the inspiration for the title of this work.
I am grateful for the hospitality of the theory groups at LBNL and UC Berkeley during the final stages of this work.
This project has received funding from the European Union’s Horizon Europe research and innovation programme
under the Marie Skłodowska-Curie Staff Exchange grant agreement No 101086085 – ASYMMETRY.
Computations for this work were performed on
the local SOM clusters at the Instituto de F\'isica Corpuscular (IFIC), including \texttt{Gluon} and
\texttt{Graviton}. I acknowledge support from a Feodor Lynen Fellowship of the Alexander von Humboldt
Foundation, and I am supported by the Spanish National Grant
PID2025-171343NB-C55 and by the Generalitat Valenciana through the grant CIPROM/22/69. I used Claude Opus 5.5
(Anthropic) for assistance in coding and runtime optimisations of \TL. I verified all of its output and take
full responsibility for the content of this work.

\appendix

\section{Bubble nucleation rate in the three-dimensional EFT}
\label{app:rate}

\begin{table}[!htbp]
\centering
\small
\begin{tabular}{lrrr}
\toprule
$g$ & 0.475 & 0.500 & 0.700 \\
\midrule
$T$ [GeV] & $3.435\times10^{-5}$ & $2.103\times10^{-4}$ & $4.682\times10^{-2}$ \\
$m_3^2$ [GeV$^2$] & $5.543\times10^{-11}$ & $2.336\times10^{-9}$ & $2.373\times10^{-4}$ \\
$\lambda_3$ [GeV] & $-5.716\times10^{-7}$ & $-3.290\times10^{-6}$ & $-3.404\times10^{-4}$ \\
$g_3^2$ [GeV] & $7.689\times10^{-6}$ & $5.221\times10^{-5}$ & $2.289\times10^{-2}$ \\
$m_{\rm D}^2$ [GeV$^2$] & $8.830\times10^{-11}$ & $3.672\times10^{-9}$ & $3.597\times10^{-4}$ \\
$\lambda_{\rm VL}$ [GeV] & $1.539\times10^{-5}$ & $1.046\times10^{-4}$ & $4.595\times10^{-2}$ \\
\midrule
$\Sthree^{\rm NLO}/T$ & $140.6415$ & $146.9802$ & $154.5403$ \\
$\delta_h$ & $1.9930$ & $2.0046$ & $3.2844$ \\
$\delta_\chi$ & $-23.0371$ & $-27.7323$ & $-69.7783$ \\
$\delta_1$ & $-25.7211$ & $-30.6143$ & $-74.3876$ \\
$\delta_0$ & $-29.9989$ & $-35.3506$ & $-83.7369$ \\
$\tfrac12\Delta_{\rm mix}$ & $-18.1732$ & $-19.2322$ & $-26.0894$ \\
$S^{J=0}$ & $0.9350$ & $0.9190$ & $0.8651$ \\
$-C$ & $83.0272$ & $98.2844$ & $234.9636$ \\
$\int V_3^{\rm NLO}$ & $2.9346$ & $3.7605$ & $13.7310$ \\
$S_6$, eq.~\eqref{eq:S6} & $0.1604$ & $0.2420$ & $1.7596$ \\
\midrule
$E$, eq.~\eqref{eq:assembly} & $132.7614$ & $139.2612$ & $155.1518$ \\
\midrule
$-\ln(\Adyn/T)$ & $2.3626$ & $2.3432$ & $2.3434$ \\
$\tfrac32\ln(S_{\rm tot}/2\pi)$ & $4.6533$ & $4.7219$ & $4.8780$ \\
\midrule
$\Delta S$, eq.~\eqref{eq:deltaS} & $-0.8643$ & $-0.6539$ & $7.8330$ \\
\bottomrule
\end{tabular}
\caption{The terms of eqs.~\eqref{eq:assembly} and~\eqref{eq:deltaS} at three couplings with $v =
\eu^{1/6}\,{\rm GeV}/g$, for the NLO matching at $\mu_4 = 7.06\,T$ and $\mu_3 = gT$, each at their percolation
temperature $T = \Tp$ obtained at NLO precision.
The interpolation of \TL\ for $\Delta S$ reproduces these values up to the fourth significant digit, way beyond the
accuracy that the PTA data requires, see discussion of variant \lk{N1}.
The values and the script that computes them are provided with the analysis layer
(\texttt{determinant\_reference\_values.csv}).
\label{tab:reference}}
\end{table}

\begin{table}[!htbp]
\centering
\footnotesize
\setlength{\tabcolsep}{3pt}
\begin{tabular}{@{}lrrrrrrrrrr@{}}
\toprule
 & & & & & & & \multicolumn{2}{c}{Theory band [\%]} & \multicolumn{2}{c}{Num.~accuracy [\%]} \\
\cmidrule(lr){8-9}\cmidrule(l){10-11}
$g$ & $\Tp$ [keV] & $\bHR$ & $\bHS$ & $\bHG$ & $\frac{1}{H}\frac{\dd\ln a^3(1-P_{\rm t})}{\dd t}$ & $\Delta S$ & $\delta\Tp$ & $\delta\bHR$ & $\delta\Tp$ & $\delta\bHR$ \\
\midrule
$0.42$ & 0.018 & 5.3 & 7.7 & 3.6 & $+1.43$ & $-0.46$ & $\pm53$ & $\pm4.9$ & $-0.04$ & $-0.19$ \\
$0.43$ & 0.142 & 6.7 & 9.1 & 5.0 & $+0.94$ & $-0.69$ & $\pm43$ & $\pm4.2$ & $-0.02$ & $-0.19$ \\
$0.44$ & 0.722 & 8.2 & 10.5 & 6.4 & $+0.44$ & $-0.85$ & $\pm38$ & $\pm3.9$ & $-0.02$ & $-0.19$ \\
$0.45$ & 2.75 & 9.6 & 11.9 & 7.8 & $-0.06$ & $-0.95$ & $\pm33$ & $\pm3.7$ & $-0.01$ & $-0.18$ \\
$0.46$ & 8.47 & 11.1 & 13.3 & 9.2 & $-0.58$ & $-1.01$ & $\pm30$ & $\pm3.6$ & $-0.01$ & $-0.16$ \\
$0.47$ & 22.2 & 12.7 & 14.8 & 10.7 & $-1.11$ & $-1.03$ & $\pm28$ & $\pm3.5$ & $-0.01$ & $-0.15$ \\
$0.475$ & 34.4 & 13.4 & 15.6 & 11.4 & $-1.36$ & $-1.03$ & $\pm27$ & $\pm3.2$ & $-0.01$ & $-0.15$ \\
$0.48$ & 51.8 & 13.9 & 16.4 & 12.1 & $-1.50$ & $-1.02$ & $\pm27$ & $\pm1.8$ & $-0.01$ & $-0.15$ \\
$0.49$ & 111 & 14.3 & 18.1 & 12.4 & $-1.65$ & $-0.97$ & $\pm24$ & $\pm3.7$ & $-0.01$ & $-0.14$ \\
$0.5$ & 210 & 16.3 & 19.8 & 14.0 & $-2.38$ & $-0.90$ & $\pm22$ & $\pm5.5$ & $0.00$ & $-0.13$ \\
$0.52$ & 619 & 21.0 & 23.2 & 18.6 & $-3.98$ & $-0.67$ & $\pm20$ & $\pm4.3$ & $0.00$ & $-0.11$ \\
$0.55$ & 2{,}191 & 27.4 & 29.1 & 24.8 & $-6.17$ & $-0.13$ & $\pm18$ & $\pm4.2$ & $0.00$ & $-0.09$ \\
$0.6$ & 9{,}387 & 39.3 & 40.8 & 36.3 & $-10.30$ & $+1.27$ & $\pm14$ & $\pm4.7$ & $0.00$ & $-0.05$ \\
$0.65$ & 24{,}447 & 51.2 & 55.3 & 48.0 & $-14.39$ & $+3.34$ & $\pm12$ & $\pm5.0$ & $0.00$ & $-0.05$ \\
$0.7$ & 46{,}822 & 67.8 & 72.8 & 63.6 & $-20.16$ & $+6.06$ & $\pm10$ & $\pm6.6$ & $0.00$ & $-0.04$ \\
$0.8$ & 102{,}979 & 106.9 & 118.7 & 101.5 & $-33.65$ & $+13.40$ & $\pm7$ & $\pm5.6$ & $+0.01$ & $-0.36$ \\
\bottomrule
\end{tabular}

\caption{Phase transition parameters for $\muz = 1$\,GeV, i.e.\ $v = \eu^{1/6}\,{\rm GeV}/g$, from the direct
integration of the true-vacuum
fraction: the percolation temperature $T_{\rm p}$, the three definitions of the phase transition speed
of eq.~\eqref{eq:speeds}, the growth rate $\dd\ln(a^3(1 - P_{\rm t}))/\dd t / H$ of the
physical false-vacuum volume at percolation, and the correction
$\Delta S$ at $\Tp$.
The following two columns list the half width of the theory band of $\Tp$ and $\bHR$, see appendix~\ref{app:rate}.
The last two columns show measures of the numerical accuracy, given by the relative difference between the full runs
of \TL\ with its adaptive percolation solver and the direct integration. \label{tab:budget_points}}
\end{table}

We compute the nucleation rate at NLO in the three-dimensional EFT using the methods presented in
refs.~\cite{Christiansen:2025xhv, Puchades-Ibanez:2026ksh}, which we
implemented in \TL. Where the choices coincide, it agrees with the independent computations of
refs.~\cite{Christiansen:2025xhv, Puchades-Ibanez:2026ksh} to $0.1\%$ in the exponent.
This brief appendix summarises the
computation, compares with the literature, and describes what our treatment adds to earlier work.
More detailed notes on the implementation and the technical
details of the NLO computation will be given in a future version~3 of \TL.

The EFT of the dark $U(1)$ and its matching relations at LO and NLO are those of
ref.~\cite{Christiansen:2025xhv}.
We generated the matching relations again with \DRalgo~\cite{Ekstedt:2022bff}, and checked that they agree term by term.
Integrating out the hard modes of order $\pi T$ leaves an EFT at the soft scale with the static scalar, with a
tree-level potential
$\tfrac12 m_3^2\phi_3^2 + \tfrac14\lambda_3\phi_3^4$, the spatial gauge field $A_i$, and the temporal gauge field
$A_0$ with the Debye mass $m_{\rm D}$.
Integrating out $A_i$ and $A_0$ at one loop in the background field $\phi_3$ results in the soft
potential~\cite{Christiansen:2025xhv, Puchades-Ibanez:2026ksh, Kierkla:2025qyz}, the tree-level potential plus
the one-loop contributions of $A_i$ and $A_0$, which are of leading order since near the bubble the gauge
field is heavier than the scalar,
\begin{equation}
  V_3 = \tfrac12 m_3^2\phi_3^2 + \tfrac14\lambda_3\phi_3^4 - \frac{1}{12\pi}\left(2m_A^3 + m_{A_0}^3 - m_{\rm D}^3\right),
  \qquad m_A^2 = g_3^2\phi_3^2\,,\quad m_{A_0}^2 = m_{\rm D}^2 + \tfrac12\lambda_{\rm VL}\phi_3^2
  \label{eq:V3}
\end{equation}
through which the critical bubble is computed.
Here, $\phi_3$ is the background field, and the mass parameter $m_3^2$, the quartic coupling $\lambda_3$, the gauge
coupling $g_3^2$, and the coupling $\lambda_{\rm VL}$ of $A_0$ to the scalar come from matching at either LO
or NLO to the four-dimensional theory.
The critical bubble $\phi_3(r)$ is the bounce of $V_3$, the spherically symmetric solution of
$\phi_3'' + (2/r)\phi_3' = \dd V_3/\dd\phi_3$ that approaches the false vacuum $\phi_3 = 0$ for $r \to \infty$.
With the parameters matched at NLO its action, dimensionless in the three-dimensional theory, is $\Sthree^{\rm NLO}/T$,
and the following corrections of eq.~\eqref{eq:assembly} are evaluated on this bubble~\cite{Ekstedt:2021kyx,
Kierkla:2025qyz}.
Since $\muz$ is the
only dimensionful parameter once $\mu_4$ and $\mu_3$ are fixed in units of $T$, the nucleation rate's exponent
depends only on $g$ and
$T/\muz$, which is of use in the numerics of \TL\ as described in appendix~\ref{app:code}.
At NLO the nucleation rate is $\Gamma = \Adyn T^3\eu^{-E}$, with~\cite{Kierkla:2025qyz, Ekstedt:2021kyx,
Bernardo:2026tki}
\begin{equation}
  E = \frac{\Sthree^{\rm NLO}}{T}
  + \delta_h + \delta_\chi + \delta_1 + \delta_0 + \tfrac12\Delta_{\rm mix} + S^{J=0} - C
  + 4\pi\int_0^\infty\!\dd r\, r^2\, V_3^{\rm NLO}\big(\phi_3(r)\big) + S_6\,.
  \label{eq:assembly}
\end{equation}
The terms $\delta_h$, $\delta_\chi$, $\delta_1$ and $\delta_0$ are the logarithms of the fluctuation determinants of
the radial mode, the Goldstone mode, the spatial gauge field, and the temporal gauge field around the bubble,
respectively, relative to the false vacuum,
in the Feynman--'t~Hooft gauge.
The term $\tfrac12\Delta_{\rm mix}$ is due to the mixing of the Goldstone mode with the gauge
field~\cite{Ekstedt:2021kyx}, and $S^{J=0}$ replaces the zero mode of a global $U(1)$ rotation of the bubble by
an integral over its orientation~\cite{Bernardo:2026tki}.
The contribution $C$ subtracts the gauge loops that the soft potential $V_3$ already contains,
since in the tail of the bubble the gauge field becomes lighter than the scalar~\cite{Gould:2021ccf,
Kierkla:2025qyz}. The integral is the two-loop potential of the EFT~\cite{Ekstedt:2021kyx,
Christiansen:2025xhv} and the last term is the action of the operators of dimension six. It includes
$c_6|\Phi|^6$ and $c_D(\Phi^\dagger\Phi)(D_i\Phi)^\dagger(D_i\Phi)$ on the bubble, with $\Phi = \phi_3/\sqrt2$ and the
on-shell coefficients $c_6$ and $c_D$ of ref.~\cite{Bernardo:2026tki} (see also ref.~\cite{Bernardo:2025vkz}),
hence\footnote{During the preparation of this work ref.~\cite{Bernardo:2026tki} appeared, recommending to
include the operators of dimension six in the fluctuation determinants.
The resulting shifts of the exponent are of the same order as that of those sextic operators alone, such that
the authors find the difference between both approaches to lie within the theoretical uncertainties of their
calculation.
As the perturbative expansion in any case gives the smallest contribution to the total modelling uncertainty
(see section~\ref{sec:alternatives}), we neglect this improvement in our work.}
\begin{equation}
  S_6 = 4\pi\int_0^\infty\!\dd r\, r^2\left[\frac{c_6}{8}\,\phi_3^6 + \frac{c_D}{4}\,\phi_3^2\left(\frac{\dd\phi_3}{\dd r}\right)^2\right] .
  \label{eq:S6}
\end{equation}

For the dynamical prefactor, we take Langer's prescription $\Adyn = \sqrt{|\lambda_-|}/2\pi$. Equating the rate with
eq.~\eqref{eq:Stot} defines the correction
\begin{equation}
  \Delta S = E - \frac{\Sthree^{\rm NLO}}{T} - \ln\frac{\Adyn}{T} + \frac32\ln\frac{S_{\rm tot}}{2\pi}\,,
  \label{eq:deltaS}
\end{equation}
whose terms table~\ref{tab:reference} lists at three couplings for reference and comparison with independent
implementations.
The determinants and $-C$ largely cancel, so that
$\Delta S$ is of order one in the PTA-preferred regions.

Refs.~\cite{Christiansen:2025xhv, Puchades-Ibanez:2026ksh} computed the action and the nucleation rate of this model
in the EFT independently of this work. \TL\ reproduces the action of the four three-dimensional schemes of
ref.~\cite{Christiansen:2025xhv} to $1\%$. From the tables underlying figures~1 and~2 of
ref.~\cite{Puchades-Ibanez:2026ksh} we reproduce the LO action, the two-loop potential, the subtraction $C$ and the
contribution of the determinants to the exponent to better than $0.1\%$. The exponents of the tables that the
authors shared with us (private communication) agree with ours to $0.1\%$. Their
matching scale $2\pi T$ instead of $7.06\,T$ hardly moves the preferred region (see our discussion of variant \lk{N1}).
With the choices matched to
those of their fits, the Hellings--Downs spectrum (\lk{B2}), $\bHR$ in the expansion factors of the template, and $30$
IPTA bins (\lk{B4}) against their $25$, the best fits agree within $0.005$ in $g$ and $8\%$ in $v$ for NANOGrav and
within $0.001$ in $g$ and $30\%$ in $v$ for IPTA. We expect these differences to be
due to numerical uncertainties and sampling effects in ref.~\cite{Puchades-Ibanez:2026ksh}, as well as differing
modelling choices in the steps of the analysis following the computation of the nucleation rates.

Our own additions concern the implementation and the checks. \TL\ computes the whole chain from the
four-dimensional couplings to $\Delta S$.
The single channels are computed with the Gelfand--Yaglom method in \texttt{BubbleDet}~\cite{Ekstedt:2023sqc},
which does not treat the coupled block of the Goldstone mode and the gauge field. For this block we wrote our own
solver of the coupled radial equations.
It sums the partial waves up to total angular momentum $J = 70$ and fits the remainder,
about one unit at $g = 0.475$, to better than $0.1$ units.
We keep the loops of the scalar in the determinants, since its masses squared in $V_3$ are negative inside the bubble.
Resumming them into $V_3$ as in ref.~\cite{Puchades-Ibanez:2026ksh} raises $\Tp$ by $0.5\%$ to $0.8\%$ and lowers
$\alpha$ by $2\%$ to $3.5\%$ for $0.5 \le g \le 0.8$.
Ref.~\cite{Puchades-Ibanez:2026ksh} estimated the sextic operator and left it out of its results; we include both
operators throughout our analysis. $S_6$ amounts to $0.16$ and $0.24$ at $g = 0.475$ and $0.5$, respectively.

The theory band of table~\ref{tab:budget_points} combines in
quadrature the variation of \textit{i)} $\mu_4$ and $\mu_3$ by a factor of two, \textit{ii)} the shift that
$S_6$ causes, as an estimate of the dimension-six terms in the fluctuation determinants, which we omit and
which ref.~\cite{Bernardo:2026tki} expects to be smaller than the shift of the bounce action, and
\textit{iii)} an estimate of
the next order, $\pm x_3|\delta S|$, with the loop expansion parameter $x_3 = g_3^2/(\pi m_3)$ of the EFT.
The theory band indicates that $T_\text{p}$ is uncertain up to $\mathcal{O}(50\%)$ for strong supercooling and
that $\bHR$ is uncertain up to $7\%$, which is the scale at which alternatives \lk{N1} and \lk{N2} move the
preferred region, which are the lowest of the whole presented analysis.
The numerical accuracy of the computations in \TL\ (see following section) are orders of magnitude better,
such that these do not limit the analysis.
We moreover checked that in the
$2\sigma$ regions of both data sets the mass of the gauge field at the centre of the bubble is at most $0.46$ times the
hard scale $2\pi T$, $x_3$ lies between $0.29$ and $0.38$, and the sextic operator shifts the exponent by less than
$0.4\%$, so that the EFT applies there, as also found in ref.~\cite{Bernardo:2026tki}.

\section{Fast phase transition predictions in TransitionListener}
\label{app:code}

The fits presented in this work require unprecedented speed in the prediction of the phase transition
parameters and of the GW spectrum to make them numerically feasible.
With the version of \TL being released with the present work, one point of
the $(g, v)$ plane, from the effective potential to the GW spectrum, takes about half a second on a single
core once the interpreter runs.
On the same core the public release v2.2.1 of \TL\ with the
four-dimensional daisy-resummed rate takes $7$ to $8$ minutes per point of the preferred regions, about $650$ to
$800$ times longer, due to the strong supercooling and the required amount of iterations over the percolation
integral in eq.~\eqref{eq:percInt} to reach the desired accuracy.
The majority of the stated runtime is spent on bounce action computations.
Predictions using the numerically much more demanding NLO precision would hence be intractable without runtime
optimisation.
The most relevant optimisations are the following:
\begin{itemize}[label=\bubblebullet, leftmargin=1.6em, itemsep=1pt, parsep=0pt, topsep=3pt]
\item \textbf{The scaling of the action:} $\Sthree/T$ depends only on $g$ and $T/\muz$
(see appendix~\ref{app:rate}), so that the rate at any other $v$ follows from a one-dimensional scan on the
line with $\muz = 1$\,GeV by rescaling the
temperature in proportion to $v$, and one table of $\Sthree/T$ per coupling holds for every $v$.
\item \textbf{Tables of the NLO correction:} The numerically expensive functional determinants are computed
once, on a grid of couplings and as a
function of the LO action. The resulting table of $\Delta S$ for this model are shipped together with \TL.
\item \textbf{Splines:} The actions and the phase transition parameters are interpolated with shape-preserving
cubic splines
in the temperature and between the couplings.
\item \textbf{Pre-compiled code:} The percolation integral, the expansion history, the root searches, and the
bounce solver are now pre-compiled, so that the interpreter does not need to parse them at every point.
\item \textbf{Direct integration:} Integrating the true-vacuum fraction on the tabulated rates, for every
variant involving changes of the nucleation rate and of the percolation computation, further reduces the
runtime.
\end{itemize}

The predictions achieved in this way are robust and precise.
Full runs of \TL, with its adaptive percolation solver without relying on cached results, reproduce $\Tp$
and $\bHR$ of the direct integration to better than $0.5\%$ (see table~\ref{tab:budget_points}) for most variants.
The variants of the
nucleation rate and of percolation change $\Treh$ by less than $1\%$, except $V_{\rm 4D}$ + Daisy at the vacuum scale
(\lk{N5}), which changes it by a few per cent, and $V_{\rm 4D}$ + Daisy at the thermal scale (\lk{N4}), which
changes it by at most $1.1\%$ for $g = 0.46$ to $0.8$. 

\TL\ is easy to use for such a study.
Within the configuration file most alternatives of table~\ref{tab:switches} can be varied and combined by setting the
appropriate keys.
The file \texttt{examples/example\_point\_3d.yaml} of the repository sets up one point: it names the model file,
the potential, the scan type \texttt{SinglePoint}, the values of $g$ and $v$, every key of table~\ref{tab:switches} with its alternatives, and the output format and directory.
The command \texttt{tl -c examples/example\_point\_3d.yaml} computes this point, from the potential to the GW spectrum
and its observability, in about ten seconds including the start of the interpreter.
All presented results can be reproduced within about six hours when running the prepared scripts on
a node with 56 cores, when using the shipped action caches. Regenerating the action caches takes
about a day on the same node.

\section{Statistical analysis}
\label{app:inference}

This appendix describes the likelihood, the priors, and the test statistic used to evaluate the preference for
the phase transition, presents the astrophysical
priors of the binaries, and compares how marginalisation and profiling can lead to different preferred regions
in the $(g, v)$ plane.

\subsection{Likelihood, priors, and test statistic}
\label{app:preference}

The likelihood $L(d\,|\,g, v, \theta)$ of the PTA data $d$ is the product over the frequency bins of the
free-spectral densities at the power of the predicted spectrum, where $\theta$ are the six
astrophysical parameters of the binaries when they are included. The priors are flat in $g$ and in $\log_{10} v$ on
$0.39 < g < 0.75$ and $10\,{\rm MeV} < v < 100\,{\rm GeV}$, and zero where the phase transition does not
complete, since there the bubbles do not collide, no GWs are emitted and the false vacuum keeps inflating.
The
astrophysical parameters follow the prior $\pi(\theta)$, see table~\ref{tab:binpriors} and
ref.~\cite{Bringmann:2026uxo} for further details.

The preference for the phase transition compares two hypotheses: the binaries alone fitting the PTA data, and
the binaries together with
the phase transition, requiring the introduction of two extra parameters $g$ and $v$, achieving the same.
For each hypothesis we take the maximum of the
logarithm of the likelihood plus that of the astrophysical prior, so that the prior acts as a penalty on populations
that the galaxy surveys disfavour. The gain is the difference of the two maxima,
\begin{equation}
  G = \max_{g,\,v,\,\theta}\left[\ln L(d\,|\,g, v, \theta) + \ln \pi(\theta)\right]
  - \max_{\theta}\left[\ln L(d\,|\,\theta) + \ln \pi(\theta)\right].
  \label{eq:gain}
\end{equation}
The binaries alone are the limit of a vanishing contribution of the phase transition, which lies on the boundary of
the larger model. We approximate the distribution of $2G$ under the binaries alone by an equal mixture of zero and a
$\chi^2$ distribution with two degrees of freedom~\cite{Chernoff:1954eli}, so that a gain at least as large as the
observed one has the probability
\begin{equation}
  p = \tfrac12\,P\big(\chi^2_2 > 2G\big) = \tfrac12\,\eu^{-G}\,,
  \label{eq:pgain}
\end{equation}
which we convert to a one-sided Gaussian significance, as in ref.~\cite{Bringmann:2026uxo}.
We note that this is an approximation, as it assumes that all parameters of the larger model are defined under
the null hypothesis, whereas $g$
and $v$ are undefined once the contribution of the phase transition vanishes. The significance is moreover local,
since it does not account for the look-elsewhere-effect of the search over the $(g, v)$ plane.

\begin{table}[t]
\centering
\small
\begin{tabular}{@{}lcccc@{}}
\toprule
Parameter & Range & NANOGrav & Tomczak--Kormendy & Galaxy survey-averaged \\
\midrule
$\log_{10}(\psi_0/{\rm Mpc}^{-3})$ & $[-3.5, -1.5]$ & $-2.56 \pm 0.40$ & $-2.61 \pm 0.10$ & $-2.20 \pm 0.15$ \\
$\log_{10}(m_{\psi,0}/M_\odot)$ & $[10.5, 12.5]$ & $10.90 \pm 0.40$ & $10.96 \pm 0.11$ & $10.82 \pm 0.11$ \\
$\log_{10}(\mu/M_\odot)$ & $[7.6, 9.0]$ & $8.60 \pm 0.20$ & $8.69 \pm 0.05$ & $8.80 \pm 0.10$ \\
$\epsilon$ [dex] & $[0, 0.9]$ & $0.32 \pm 0.15$ & $0.28 \pm 0.05$ & $0.40 \pm 0.05$ \\
$\tau_{\rm f}$ [Gyr] & $[0.1, 11]$ & Uniform & Uniform & Uniform \\
$\nu_{\rm inner}$ & $[-1.5, 0]$ & Uniform & Uniform & Uniform \\
\bottomrule
\end{tabular}
\caption{The six astrophysical parameters of the binary population described in section~\ref{sec:data} and their three
prior sets: Gaussians with the given mean and standard deviation, truncated to the range of the \texttt{holodeck}
library on which the emulator is fitted, and uniform in $\tau_{\rm f}$ (in Gyr, not in its logarithm) and in
$\nu_{\rm inner}$.
$\mu$ is the black
hole mass for a bulge mass of $10^{11}\,M_\odot$, and $\epsilon$ the scatter of the black hole
mass around that value, measured in dex~\cite{Bringmann:2026uxo, NANOGrav:2023hfp}.\label{tab:binpriors}}
\end{table}

\subsection{Marginalisation and profiling of the likelihood}
\label{app:profiling}

\begin{figure}[t]
\centering
\includegraphics[width=0.85\textwidth]{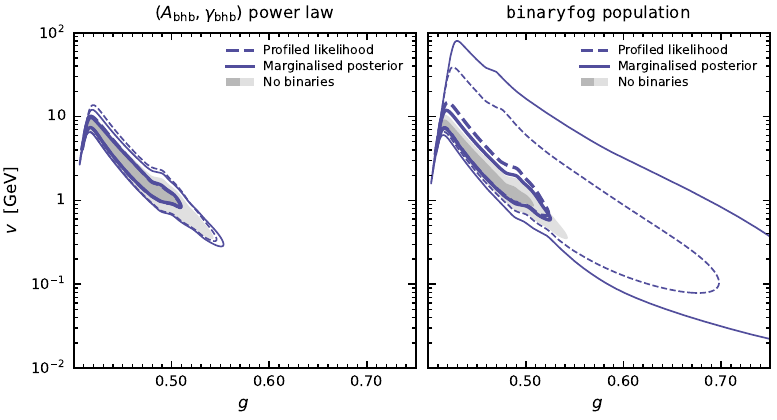}
\caption{The NANOGrav $1\sigma$ (thick) and $2\sigma$ (thin) regions with the binary background, for the
$(A_{\rm bhb}, \gamma_{\rm bhb})$ power law (left) and the \texttt{binaryfog} population with the
survey-averaged priors (right).
Solid lines indicate the posterior with the binary background marginalised over its prior.
Dashed lines instead depict contours of the profiled likelihood, with the
background parameters maximised at every point. In the latter case the
$1\sigma$ and $2\sigma$ contours correspond to $\Delta\ln L = 1.15$ and $3.09$
below the maximum. With
the binary population the $2\sigma$ profiled region closes at $g \simeq 0.70$, whereas the marginalised
$2\sigma$ region reaches the end of the prior range at $g = 0.75$.
Grey contours show the posterior without binaries.
\label{fig:foreground}}
\end{figure}

Two questions can be asked of the likelihood with the binaries, and they are answered by two different treatments of
the astrophysical parameters.
The first question is where in the $(g, v)$ plane the phase transition is most plausible, given what is known
about the binaries. It is answered by the posterior with the astrophysical parameters marginalised over their prior:
populations that the galaxy surveys disfavour enter with a small weight, and the many populations that explain the
signal without the phase transition give weight to the points where the phase transition contributes little,
such that the phase transition contribution has to make up for the difference.
This is
the treatment of figures~\ref{fig:plane} and~\ref{fig:budget}. The second question is how much better the data are
fitted by the combination of a phase transition signal and a binary signal than by the best population model.
It is answered by the profiled likelihood, with the
astrophysical parameters maximised at every point, and by the gain of eq.~\eqref{eq:gain}.

Figure~\ref{fig:foreground} compares both treatments. With the power law, the two treatments give nearly the same
regions. With the population, both $1\sigma$ regions end at $g \simeq 0.52$,
since the binaries alone fit the data worse than the best fit with the phase transition. The $2\sigma$ regions open
towards large couplings, where the binaries produce most of the signal and the phase transition contributes
only a small part of
it: the marginalised one up to the end of the prior range, $g = 0.75$, the profiled one up to $g \simeq 0.70$.
Both treatments are consistent with each other and imply that at the $2\sigma$ level (and with flat priors for
the phenomenological hardening parameters), the data constrains the phase transition and the binary population
hardly beyond their priors, apart from excluding populations that would exceed the measured power.
We explicitly checked the latter statement by repeating the fit with an MCMC chain and comparing prior and
posterior distributions of the astrophysical parameters.

\section{Modelling choices with the IPTA DR2 data}
\label{app:ipta}

For the IPTA DR2 data we repeat the comparison of section~\ref{sec:alternatives}.
The binary population is emulated again
on the IPTA frequencies, which start at $1.05$\,nHz, below the NANOGrav band: the same \texttt{holodeck} models as in
ref.~\cite{Bringmann:2026uxo} are computed on these frequencies, and the distribution of the background is
fitted as there.

\begin{table}[!tp]
\centering
\footnotesize
\setlength{\tabcolsep}{2.1pt}
\begin{tabular}{@{}llllllrrr@{}}
\toprule
 & & \multicolumn{2}{c}{$g$} & \multicolumn{2}{c}{$v$ [GeV]} & & & \\
\cmidrule(lr){3-4}\cmidrule(lr){5-6}
 & & $1\sigma$ & $2\sigma$ & $1\sigma$ & $2\sigma$ & $\mathcal{D}$ & Area & $\Delta\ln L$ \\
\midrule
\multicolumn{2}{@{}l}{This work's default} & 0.433--0.497 & 0.414--0.532 & 0.96--4.8 & 0.43--7.9 & & & \\
\midrule
\lk{N1} & $\mu_4/2$ or $2\mu_4$ & 0.433--0.497 & 0.414--0.532 & 0.96--4.8 & 0.43--7.9 & 0.0 & 1.00 & $0.00$ \\
\lk{N1} & $\mu_3/2$ or $2\mu_3$ & 0.433--0.497 & 0.413--0.531 & 0.95--4.8 & 0.43--7.9 & 0.1 & 1.00 & $0.00$ \\
\lk{N2} & $A_{\rm dyn}\times10$ or $/10$ & 0.436--0.500 & 0.417--0.536 & 0.97--4.7 & 0.42--7.7 & 0.4 & 1.00 & $0.00$ \\
\lk{N3} & LO instead of NLO & 0.439--0.506 & 0.418--0.544 & 0.92--4.7 & 0.41--7.7 & 0.9 & 1.05 & $0.00$ \\
\lk{N4} & $V_{\rm 4D}$ + Daisy at the thermal scale & 0.438--0.504 & 0.417--0.542 & 0.96--4.7 & 0.41--7.7 & 0.8 & 1.03 & $0.00$ \\
\lk{N5} & $V_{\rm 4D}$ + Daisy at the vacuum scale & 0.483--0.541 & 0.463--0.581 & 0.97--4.0 & 0.37--6.1 & 5.9 & 0.90 & $0.00$ \\
\midrule
\lk{P1} & $H^2 \approx \rho_{\rm f}/3M_{\rm Pl}^2$ & 0.437--0.499 & 0.415--0.534 & 0.99--4.7 & 0.42--8.7 & 0.4 & 1.00 & $0.00$ \\
\lk{P2} & $f_{\rm perc} = 0.63$ instead of $0.29$ & 0.430--0.496 & 0.414--0.531 & 0.92--4.6 & 0.43--6.8 & 0.4 & 0.99 & $0.00$ \\
\lk{P3} & $a^3(1-P_{\rm t})$ decreases at percolation & 0.457--0.505 & 0.449--0.539 & 0.81--2.3 & 0.36--3.0 & 0.5 & 0.74 & $0.00$ \\
\lk{P4} & $(\beta/H)_S$ instead of $(\beta/H)_R$ and $(\beta/H)_\Gamma$ & 0.425--0.484 & 0.411--0.525 & 0.89--4.3 & 0.37--7.9 & 1.6 & 0.88 & $0.00$ \\
\lk{P5} & Instantaneous reheating & 0.430--0.501 & 0.410--0.535 & 0.69--2.7 & 0.38--4.0 & 2.2 & 0.93 & $0.00$ \\
\midrule
\lk{G1} & No expansion between bubbles & 0.453--0.506 & 0.437--0.541 & 0.97--3.5 & 0.38--6.2 & 1.2 & 0.87 & $-0.09$ \\
\lk{G2} & Free-streaming shells & 0.428--0.490 & 0.414--0.530 & 1.8--8.0 & 0.62--11 & 2.4 & 0.96 & $-0.21$ \\
\lk{G3} & Higgsless template (LISA 2024) & 0.504--0.601 & 0.460--0.644 & 0.10--0.77 & 0.05--2.0 & 2.9 & 1.38 & $-0.04$ \\
\lk{G4} & Hybrid template (LISA 2019) & 0.539--0.622 & 0.492--0.668 & 0.04--0.16 & 0.02--0.35 & 4.6 & 1.01 & $-0.38$ \\
\lk{G5} & Envelope approximation & 0.442--0.502 & 0.426--0.535 & 0.31--1.7 & 0.17--3.7 & 4.7 & 0.88 & $-0.41$ \\
\midrule
\lk{B1} & $(A, \gamma)$ power-law binaries & 0.432--0.499 & 0.414--0.543 & 0.98--5.2 & 0.39--8.5 & 0.4 & 1.12 & $-0.01$ \\
\lk{B1} & Survey-avg. population prior & 0.433--0.525 & 0.414--0.689 & 0.83--6.4 & 0.15--18 & 2.9 & 2.01 & $-0.02$ \\
\lk{B1} & NANOGrav population prior & 0.433--0.510 & 0.414--0.665 & 0.91--5.7 & 0.23--12 & 1.9 & 1.47 & $0.00$ \\
\lk{B1} & Tomczak--Kormendy population prior & 0.432--0.499 & 0.414--0.541 & 0.97--5.2 & 0.38--8.4 & 0.3 & 1.14 & $0.00$ \\
\lk{B2} & Hellings--Downs instead of CURN & 0.437--0.505 & 0.415--0.544 & 0.97--4.8 & 0.38--8.6 & 0.8 & 1.13 & -- \\
\lk{B3} & DMGP instead of DMX & 0.483--0.577 & 0.447--0.611 & 0.18--1.6 & 0.09--3.7 & 2.5 & 1.92 & -- \\
\lk{B4} & 5 bins & 0.455--0.552 & 0.419--0.587 & 0.26--2.6 & 0.13--6.9 & 1.2 & 1.83 & -- \\
\lk{B4} & 30 bins & 0.430--0.495 & 0.413--0.529 & 1.0--5.2 & 0.46--8.1 & 0.1 & 0.96 & -- \\
\lk{B5} & IPTA DR2 instead of NANOGrav 15$\,$yr & 0.470--0.519 & 0.445--0.547 & 0.24--0.70 & 0.13--1.5 & 5.0 & 1.52 & -- \\
\bottomrule
\end{tabular}

\caption{The alternatives of figure~\ref{fig:budget} for the NANOGrav 15\,yr data: the marginal
$1\sigma$ and $2\sigma$ intervals of $g$ and of $v$, the distance $\DM$ of eq.~\eqref{eq:distance}
between the posterior means, the ratio of the $1\sigma$ areas and the change of the maximum of
$\ln L$ with respect to the default, given only where the likelihood function is the same as for the
default or, for \lk{B1}, extends it by the binary background.
For \lk{B1} the column gives $\ln L$ without the prior term $\ln\pi$ of the binaries at the maximum of $\ln L
+ \ln\pi$ over $(g, v)$ and the parameters of the binaries.
No interval reaches the edge of the prior range, $0.39 < g < 0.75$ and $10\,{\rm MeV} < v < 100\,{\rm GeV}$;
the intervals of \lk{B1}, however, depend on this range (see section~\ref{sec:alt_data}).
For \lk{N1} and \lk{N2} the table gives the direction with the larger $\DM$.
\label{tab:switch_numbers}}
\end{table}

\begin{table}[!tp]
\centering
\footnotesize
\setlength{\tabcolsep}{2.1pt}
\begin{tabular}{@{}llllllrrr@{}}
\toprule
 & & \multicolumn{2}{c}{$g$} & \multicolumn{2}{c}{$v$ [GeV]} & & & \\
\cmidrule(lr){3-4}\cmidrule(lr){5-6}
 & & $1\sigma$ & $2\sigma$ & $1\sigma$ & $2\sigma$ & $\mathcal{D}$ & Area & $\Delta\ln L$ \\
\midrule
\multicolumn{2}{@{}l}{This work's default} & 0.470--0.519 & 0.445--0.547 & 0.24--0.70 & 0.13--1.5 & & & \\
\midrule
\lk{N1} & $\mu_4/2$ or $2\mu_4$ & 0.470--0.519 & 0.445--0.546 & 0.24--0.70 & 0.13--1.6 & 0.0 & 1.00 & $0.00$ \\
\lk{N1} & $\mu_3/2$ or $2\mu_3$ & 0.469--0.519 & 0.444--0.546 & 0.24--0.71 & 0.13--1.5 & 0.0 & 1.00 & $0.00$ \\
\lk{N2} & $A_{\rm dyn}\times10$ or $/10$ & 0.473--0.522 & 0.448--0.550 & 0.24--0.70 & 0.13--1.5 & 0.2 & 1.00 & $0.00$ \\
\lk{N3} & LO instead of NLO & 0.478--0.530 & 0.451--0.559 & 0.23--0.69 & 0.13--1.5 & 0.7 & 1.05 & $0.00$ \\
\lk{N4} & $V_{\rm 4D}$ + Daisy at the thermal scale & 0.477--0.528 & 0.451--0.556 & 0.23--0.69 & 0.13--1.5 & 0.6 & 1.03 & $0.00$ \\
\lk{N5} & $V_{\rm 4D}$ + Daisy at the vacuum scale & 0.518--0.563 & 0.493--0.593 & 0.23--0.68 & 0.12--1.4 & 3.4 & 0.95 & $+0.05$ \\
\midrule
\lk{P1} & $H^2 \approx \rho_{\rm f}/3M_{\rm Pl}^2$ & 0.473--0.521 & 0.449--0.548 & 0.24--0.70 & 0.13--1.5 & 0.2 & 0.99 & $0.00$ \\
\lk{P2} & $f_{\rm perc} = 0.63$ instead of $0.29$ & 0.467--0.518 & 0.440--0.546 & 0.24--0.72 & 0.13--1.6 & 0.2 & 1.03 & $0.00$ \\
\lk{P3} & $a^3(1-P_{\rm t})$ decreases at percolation & 0.471--0.520 & 0.453--0.547 & 0.24--0.67 & 0.13--1.2 & 0.1 & 0.96 & $0.00$ \\
\lk{P4} & $(\beta/H)_S$ instead of $(\beta/H)_R$ and $(\beta/H)_\Gamma$ & 0.456--0.502 & 0.434--0.536 & 0.24--0.76 & 0.13--1.5 & 1.0 & 0.92 & $0.00$ \\
\lk{P5} & Instantaneous reheating & 0.469--0.521 & 0.444--0.549 & 0.20--0.54 & 0.12--1.1 & 0.6 & 0.98 & $0.00$ \\
\midrule
\lk{G1} & No expansion between bubbles & 0.478--0.524 & 0.458--0.553 & 0.23--0.73 & 0.12--1.4 & 0.5 & 1.04 & $-0.19$ \\
\lk{G2} & Free-streaming shells & 0.462--0.520 & 0.431--0.552 & 0.28--1.5 & 0.14--3.7 & 1.3 & 1.50 & $-0.53$ \\
\lk{G3} & Higgsless template (LISA 2024) & 0.575--0.626 & 0.545--0.658 & 0.03--0.08 & 0.02--0.16 & 4.0 & 0.82 & $+0.13$ \\
\lk{G4} & Hybrid template (LISA 2019) & 0.589--0.639 & 0.563--0.664 & 0.01--0.03 & 0.01--0.05 & 4.9 & 0.63 & $+0.60$ \\
\lk{G5} & Envelope approximation & 0.462--0.502 & 0.447--0.540 & 0.12--0.37 & 0.06--0.63 & 2.9 & 0.72 & $+0.21$ \\
\midrule
\lk{B1} & $(A, \gamma)$ power-law binaries & 0.472--0.537 & 0.443--0.684 & 0.21--0.85 & 0.07--14 & 1.5 & 1.60 & $-0.08$ \\
\lk{B1} & Survey-avg. population prior & 0.474--0.561 & 0.442--0.711 & 0.19--1.0 & 0.04--32 & 2.5 & 2.22 & $-0.02$ \\
\lk{B1} & NANOGrav population prior & 0.473--0.587 & 0.441--0.723 & 0.20--1.5 & 0.04--46 & 3.3 & 2.76 & $0.00$ \\
\lk{B1} & Tomczak--Kormendy population prior & 0.471--0.522 & 0.445--0.554 & 0.23--0.71 & 0.12--1.6 & 0.1 & 1.09 & $-0.01$ \\
\lk{B4} & 5 bins & 0.430--0.514 & 0.411--0.552 & 0.26--2.3 & 0.11--4.2 & 1.0 & 1.33 & -- \\
\lk{B4} & 30 bins & 0.442--0.471 & 0.411--0.493 & 0.82--1.9 & 0.49--4.0 & 1.8 & 0.42 & -- \\
\bottomrule
\end{tabular}

\caption{The alternatives for the IPTA DR2 data, as table~\ref{tab:switch_numbers} for NANOGrav. The binaries are
evaluated through \texttt{binaryfog} on the IPTA free spectrum and compared with the IPTA likelihood without them,
evaluated in the same way.
\lk{B2}, \lk{B3} and \lk{B5} have no IPTA counterpart.
For \lk{G4} the prior range cuts off the posterior at its lower end, $v = 10$\,MeV, where the marginal
posterior of $v$ is still $12\%$ of its maximum at $v = 20$\,MeV, so that a wider prior range would move the
lower end of the $2\sigma$ interval of $v$ below $10$\,MeV, where the reheating temperature lies below the BBN
bound.\label{tab:switch_numbers_ipta}}
\end{table}

Table~\ref{tab:switch_numbers_ipta} lists the shifts of the IPTA region, to be compared with
table~\ref{tab:switch_numbers} for NANOGrav. Qualitatively, the IPTA data lead to the same conclusions as the NANOGrav
data. The perturbative uncertainties of the nucleation rate and $V_{\rm 4D}$ + Daisy with running couplings (\lk{N4})
leave the region in place, and $V_{\rm 4D}$ + Daisy at the vacuum scale (\lk{N5}) again favours couplings about
$10\%$ larger, since it makes the phase transition slower at every point of the $(g, v)$ plane.
Instantaneous reheating again lowers the
inferred $v$, and the slope $\bHS$ again lowers the inferred $g$. The spectra sourced by sound waves move the region
furthest, to faster phase transitions at a $v$ so small that the reheating temperature falls below its lower bound of
$6$\,MeV in most of the posterior. The binaries again widen the region towards large couplings, by more than for
NANOGrav, since the IPTA data alone favours the phase transition less. One alternative acts differently. With $30$
instead of $13$ frequency bins (\lk{B4}) the region narrows and moves to slower phase transitions at larger $v$,
whereas for NANOGrav the additional bins leave it in place.
For IPTA the bins above the default $13$ lie between $15$ and $33$\,nHz, on the falling side of the spectrum,
and most of them carry weak power at the level of its tail.
The fit raises the tail through a slower phase transition at a higher reheating temperature, so that the region
moves to larger $v$.
No single bin causes the shift, which builds up as the bins are added one by one.
For NANOGrav the default $14$ bins already reach beyond the peak, and the bins added above $30$\,nHz act as upper
limits far above the predicted spectrum.

\FloatBarrier

\bibliographystyle{JHEP_improved}
\bibliography{refs}

\end{document}